\documentclass[thmsa,a4paper,11pt]{article}
\usepackage{amssymb}

\usepackage{sw20lart}

\input{tcilatex}

\begin{document}

\author{M.\ Holzmann, P.\ Gr\"{u}ter\thanks{%
Present address: McKinsey \& Company, Kurf\"{u}rstendamm 185, D-10707
Berlin, Germany.} \thinspace and F.\ Lalo\"{e} \\
%EndAName
\\
LKB\thanks{%
The Laboratoire Kastler Brossel is Unit\'{e} Associ\'{e}e au CNRS (UA 18) et
\`{a} l'Universit\'{e} Pierre et Marie Curie.} and LPS\thanks{%
The Laboratoire de Physique Statistique de l'ENS is Unit\'{e} Associ\'{e}e
au CNRS (UA 1306) et aux Universit\'{e}s Paris 6 et Paris 7.} ,
D\'{e}partement de Physique de l'ENS\\
24 rue Lhomond, F 75005 Paris, France\\
\\
and (M.H. and F.L.)\\
Institute of Theoretical Physics,UCSB\\
Santa Barbara Cal. 93106, USA}
\title{Bose-Einstein condensation in interacting gases }
\date{}
\maketitle

\begin{abstract}
We study the occurrence of a Bose-Einstein transition in a dilute gas with
repulsive interactions, starting from temperatures above the transition
temperature.\ The formalism, based on the use of Ursell operators, allows us
to evaluate the one-particle density operator with more flexibility than in
mean-field theories, since it does not necessarily coincide with that of an
ideal gas with adjustable parameters (chemical potential, etc.). In a first
step, a simple approximation is used (Ursell-Dyson approximation), which
allow us to recover results which are similar to those of the usual
mean-field theories.\ In a second step, a more precise treatment of the
correlations and velocity dependence of the populations in the system is
elaborated. This introduces new physical effects, such as a marked change of
the velocity profile just above the transition: low velocities are more
populated than in an ideal gas.\ A consequence of this distortion is an
increase of the critical temperature (at constant density) of the Bose gas,
in agreement with those of recent path integral Monte-Carlo calculations for
hard spheres.
\end{abstract}

\ 

\section{Introduction}

The notion of Bose-Einstein condensation is not new: it was introduced by
A.\ Einstein in 1925 \cite{Einstein}; Bose himself played an important role
in the introduction of Bose-Einstein statistics, but his work was focussed
on radiation (photons) - he did not generalize it to massive particles and
therefore played no role in the discovery of the phase transition \cite
{Blanpied}\cite{Theimer}.\ It is well known that, for many years, the so
called ``Einstein phenomenon''\cite{Lamb} was considered more as a
mathematical artifact of the formalism than a physical reality - this was
for instance the case of Uhlenbeck himself \cite{Pais} \cite{AE}\cite
{Uhlenbeck}. This reaction is perfectly natural: indeed, the notion of
accumulating particles into one single quantum state - with no limit on the
accuracy of the definition of their momentum, except the size of the
macroscopic container - looks rather paradoxical at first sight.\ For an
ideal gas, it certainly introduces unphysical properties: density profiles
which depend critically on the boundary conditions of the wave functions on
the walls, in violation of extensivity \cite{Balian-BEC}, or anomalous
fluctuations of particle numbers, which become ensemble dependent \cite
{Kleppner-Greytak}\cite{Balian2}\cite{Politzer}. It is now well understood
that these pathological features disappear as soon as some repulsive
interaction between the particles is added; but then, by naive analogy, one
could ask why the occupancy of a single quantum state does not disappear as
well?\ After all, it would also seem perfectly natural to assume that, in an
interacting system, some finite momentum band is highly populated\footnote{%
An elementary idea, for instance, would be to suggest that the width of the
band is related by some inverse Fourier relation to the mean-free-path in
the gas.}.\ How can we show, from ab initio arguments, that the accumulation
of a finite proportion of particles into a single quantum state is indeed a
robust property against the presence of interactions?

Curiously, in view of the importance of the phenomenon, the literature
contains relatively little discussion of this question and of the
possibility of what has sometimes be called ``fractioned'' or ``smeared''
boson condensates; in particular, following London's historical intuition 
\cite{London}, most textbooks prefer to simply assume that one single state
is populated macroscopically, and then proceed to study the interesting
consequences of this Ansatz. At zero temperature, a general argument was
nevertheless given by Penrose and Onsager\footnote{%
Penrose and Onsager were the first to define the phenomenon of Bose-Einstein
condensation in an interacting system in terms of the eigenvalues of the one
particle reduced density operator; in addition, from a variational argument
concerning the ground state wave function, they give an estimation of the
condensed fraction in superfluid helium four.} in 1956 \cite{Penrose-Onsager}%
; other famous references are the work of Beliaev \cite{Belyaev}, of Yang 
\cite{Yang} and of Fr\"{o}hlich \cite{Fröhlich}; in 1962, Girardeau \cite
{Girardeau} discussed in detail the possibility of what he called a
``generalized condensation, where no one single-particle state is
macroscopically occupied''; a more recent discussion was given by
Nozi\`{e}res \cite{Nozieres}, who concluded from a Hartree-Fock calculation
at zero temperature that repulsive interactions tend to stabilize the single
state occupancy. At finite temperatures, ``smeared condensation'' was
discussed explicitly in an article by Luban in 1962 \cite{Luban} and, more
recently and in a mean field context, by Van den Berg et al. \cite{Berg}.
There is also a large amount of beautiful numerical work on the subject \cite
{Ceperley RMP}, which leads to an impressive agreement with experimental
data, in particular in liquid helium four; but one should keep in mind that
numerical methods are subject to limitations due to finite size effects in
the evaluation of narrow peaks in occupation numbers. For a critical
discussion of experimental evidence for a condensate in superfluid helium
four, see for instance Sokol \cite{Sokol}

Two general remarks may come to mind at this stage.\ The first is that it is
known, from very general considerations, that the Bose-Einstein statistics
plays no role whatsoever at zero temperature (Boltzmann and Bose Einstein
systems have exactly the same ground state); zero temperature arguments
therefore do not directly address the question of how the statistics is able
to stabilize the single state occupancy against finite temperature
excitations. The second is that mean-field (Hartree-Fock) arguments are
based on a approximation where the system is considered as equivalent to a
gas of independent particles with modified energies, which automatically
preserves the main properties of the ideal gas (for which the occurrence of
Bose-Einstein condensation depends only on the density of states at the
origin); they do not really address the question of the robustness of single
state occupancy against all kinds of correlation that can be created in a
system by interactions, but rather assume it. A theoretical study where the
property in question would not be an ingredient, but a consequence of the
results of the calculations, could therefore be useful.

There are also other issues which are not completely settled, such as the
effects of the interactions on the transition temperature itself.\ For
instance one may ask if repulsive hard cores with short range will tend to
increase or to decrease the critical temperature (at constant density%
\footnote{%
Here, we discuss only homogeneous systems (a gas in a box), where
translation invariance ensures that the number density $n$ remains
constant.\ When this symmetry is not fullfilled, as is the case in atomic
traps, repulsive interactions have important secondary effects, in
particular, the spatial density of the system is changed\cite{Goldman}, with
a significant decrease of the number density $n$ of the atoms at the center
of the trap (at constant total number of atoms); this reduces the degeneracy
parameter $n\lambda ^{3}$at the center of the trap and, obviously, the
transition temperature as well.\ Mean-field theories can be used to account
qualitatively for the changes of spatial distributions of atomic gases in
traps and, by the same token, for this change of critical temperature.
\par
By contrast, the effects we are interested in in the present article are of
a different nature; they originate from the microscopic correlations
introduced between the particles by the iteractions, so that they remain
essentally beyond the scope of mean-field approaches.\ While the latter
always predict that the transition occurs when $n\lambda ^{3}$ is equal to
2.612.. at the point of maximum density (center of the trap), exactly as for
an ideal gas, the purpose of our study is precisely to study the changes of
this value under the effects of interactions; in other words, we are
interested in changes of the critical value of the degeneracy parameter $%
n\lambda ^{3}$, not in changes in $n$. Under these conditions, it is more
convenient to assume that the gas is contained in a a box, since then $n$
automatically remains constant, so that the pure correlations effects are
seen with no background due to density changes.}).\ In 1957, the method of
pseudopotentials \cite{Huang-1} \cite{Huang-et-coll.} was applied by Huang
and coll. to the study of the properties of the phase transition in Bose
hard spheres.\ The result is that the critical temperature should be
slightly increased (by an amount proportional to the diameter $a^{3/2}$,
where $a$ is the diameter of the spheres); see also ref. \cite{Huang-livre}%
.\ The physical interpretation given by the authors is that, by a Heisenberg
type relation, ``a spatial repulsion gives rise to a momentum space
attraction'', therefore facilitating the appearance of a condensate. But,
since then, other methods of approach to the problem have been proposed, in
particular a Hartree-Fock type approximation \cite{Fetter-Walecka}.\ It
turns out that this method provides the opposite result: a purely repulsive
potential is predicted to lower the transition temperature!\ More precisely,
it predicts that the change in $T_{c}$ depends essentially on the range of
the potential (it vanishes for a contact potential) and that, for most of
repulsive potentials with finite range (a ``top hat potential'' for
instance) the exchange interaction tends to increase the effective mass of
the particles and therefore to lower $T_{c}$. Of course, this does not
necessarily means that the two results are contradictory: each of them might
be correct in a different domain\footnote{%
One would expect the Hartree-Fock calculation to be better for dense
systems, as all mean-field theories; in these systems each atom interacts
simultaneoulsy with many others, so that it can average most of its short
range correlations with its neighbours, and experience only a mean field
from them.\ Indeed, in a dense system such as superfluid helium 4, the basic
prediction is the Hartree Fock calculation is borne out by experiments,
since the critical temperature is found to be a decreasing function of
pressure.
\par
Nevertheless the domains of validity of the two methods should overlap at
least partially (for a potential which would be at the same time weak and
with short range) so that the prediction of what shoud happen in this case
remains unclear.}.\ Later, a renormalization group calculation was given by
Toyoda \cite{Toyoda}, which also predicted a decrease of the critical
temperature; but subsequent more refined calculations by Stoof and coll.\
led to a prediction of an increase of the temperature, proportional either
to $a$ \cite{Stoof1} or to $a^{1/2}$ \cite{Stoof2}. The most recent result
was obtained by a numerical calculation based on the path integral quantum
Monte-Carlo method \cite{Gruter}; it actually predicts the existence of a
crossover between two regimes, at low and high densities, while at low
densities the critical temperature is indeed increased - but even in this
region the results do not really agree with any of the analytical
calculations mentioned above - see also ref. \cite{Markus} for another
discussion of the low density regime. One can summarize the situation by
saying that there is no present consensus on what is the theoretical
expression of the second virial correction to the Bose-Einstein transition
temperature in a dilute gas.

More generally, one can be interested in a better understanding of the
correlation that are implied by superfluidity in a gas: on a microscopic
scale, what kind of organization in both momentum and ordinary space is
responsible for the occurrence of superfluidity?\ It is generally considered
that superfluidity is an inherently different physical phenomenon from
Bose-Einstein condensation; but what is exactly the difference between the
microscopic mechanisms that lead to each of these transitions, and to what
extent should they always appear at the same time?\ Another related question
is how the macroscopic wave function and its non-linear evolution \cite
{Pitaevskii}\cite{Gross} build up from individual correlated particles, in a
dilute gas where atoms are free most of the time, collisions are binary and
short, so that mean-field methods are not necessarily well suited? The hope
is that the study of Bose-Einstein condensation in dilute gases, where more
precise ab initio calculations should be feasible than in dense systems,
will allow a better understanding of the original phenomena observed in
superfluid systems, including the stability against dissipation, vortices,
various modes of oscillation (second sound), etc.; for a general review of
the many recent theoretical contributions in the subject, see ref. \cite
{Sandro-RMP}

In this article, we will focus ourselves on the study of the effects of
binary interactions on the transition temperature, in an approach of the
problem where the populations of various individual states are kept free to
vary in any way; they may adapt to the interactions and become more and more
different from a Bose-Einstein distribution when the temperature of a gas is
progressively cooled down. For this purpose, we will use the method of
Ursell operators, already discussed in previous articles \cite{Ursell
operators}, which is well adapted to a detailed treatment of binary short
range correlations between particles due for instance to hard cores; for a
preliminary report of the method, see \cite{Trento}. Here, instead of
calculating the partition function of the system as in previous work, we
will find it more convenient to directly evaluate the one-particle density
operator; it turns out that this is not only more direct physically, but
also mathematically more convenient, mostly because the weights of the
diagrams are much simpler than those in the partition function.

It is well known \cite{Feynman}\cite{Elser}\cite{Werner} - and this was
already emphasized in \cite{Trento} in the context of Ursell operators -
that, near the transition point, larger and larger exchange cycles of
identical atoms become more and more important: infinite summations over
many sizes of diagrams are therefore necessary.\ This task can be performed
by using implicit integral equations; in a first step, we will use the
simplest form for this equation, which is reminiscent of the Dyson equation,
and this will lead us to a first form of the Ursell theory that is very
similar to a mean-field theory.\ Not surprisingly then, we will find at this
stage that the critical degeneracy parameter $n\lambda ^{3}$ is unchanged,
keeping exactly the same value as for an ideal gas; moreover, the velocity
distribution of atoms will also remain the same as for an ideal gas, with of
course some change of the effective chemical potential introduced by the
interactions. This first step is to be seen mostly as a starting base where
known results are recovered.\ The second step will involve a slightly more
elaborate integral equation for the density operator, which fortunately is
not much more complicated than the Ursell-Dyson equation so that it can be
solved by a similar method.\ The major new feature introduced is a velocity
dependence that is no longer that of an ideal gas: it includes a distortion
of the velocity profile, especially at low velocities, a feature which is
essentially beyond mean-field approximations.\ We will interpret this result
as due to a spatial re-arrangement of atoms with low velocities, which
allows them to minimize repulsion and, so to say, come closer to
Bose-Einstein condensation than the other atoms.\ Consequently, the system
has stronger tendency to populate lower states than it would have in the
absence of interactions, which favors even more the ground state, so that it
is still true that a single quantum state tends to be macroscopically
populated; in a way, our reasoning can be seen as a more elaborate version
of the argument concerning the role of interactions developed earlier by
Nozi\`{e}res \cite{Nozieres}, as we discuss in more detail below.\ An
important consequence of this effect is that the critical degeneracy
parameter is reduced (still assuming repulsive interactions), by an amount
which is compatible with the numerical Monte-Carlo results of ref. \cite
{Gruter}. More generally, this study provides a microscopic mechanism for
the approach to Bose-Einstein condensation in a dilute gas.

Section 2 below is the most technical of this article, with diagrams,
counting, etc..; it can be skipped if the reader who is prepared to accept
without immediate justification the integral equations proposed in section 3
for the one-particle density operator .

\section{Ursell expansion of the one-body density operator}

In \cite{Ursell operators} we have shown how a finite truncation of the
Ursell operator series of the grand potential can provide virial corrections
for quantum gases, even if they are partially degenerate; in \cite{Ursell
operators 2} the same technique provided simple expressions for the one and
two particle density operator. Nevertheless, in this article, we can not
limit ourselves to these results, since they do not remain valid close to
the Bose-Einstein condensation point.\ In fact any finite truncated
expression, if taken seriously, would merely exclude the phase transition,
as discussed in \S\ 3.3 of ref.\ \cite{Trento}; the transition would be
replaced by a sharp but continuous crossover phenomenon, occurring over a
finite range of parameters (which is independent of the size of the system
but becomes narrower and narrower when the density of the gas decreases).\
But this conclusion arises from an incorrect simplification: it is clear
that the validity of any truncation always ends up breaking down at some
point when the system comes sufficiently close to Bose-Einstein
condensation. Mathematically, the reason is that higher order terms in the
series of perturbations, if they have smaller coefficients in a dilute
system, also contain more denominators which diverge when the chemical
potential tends to zero\footnote{%
In Ursell diagrams, every horizontal line introduces, after a summation over
the size of the cycles, a factor $(1+f_{1})$ (where $f_{1}$ is the one
particle distribution of the ideal gas) which diverges when $\mu \rightarrow
0$. Therefore, the more lines the diagram contains, the more divergent it is.%
}; this implies stronger divergences near the transition point, so that
higher order terms always become dominant at some point. Physically, the
reason of this behavior is a divergence of the sizes of typical exchange
cycles at the transition point \cite{Feynman}\cite{Elser}.

Of course, observing divergences in perturbation series at a phase
transition point is very common in physics. In the case of Bose-Einstein
transition, nevertheless, an unusual feature is that the transition already
exists in the ideal gas, and moreover already contains a strong singularity:
the slope of the curve giving the density as a function of the chemical
potential $\mu $ becomes infinite at the Bose Einstein critical point $\mu
=0 $ (see figure 1) .\FRAME{dtbpFU}{4.3284in}{2.2589in}{0pt}{\Qcb{Fig.\ 1:
Density of an ideal gas as a function of the chemical potential $\protect\mu 
$. If the effect of the interactions was simply to shift the critical value
of $\protect\mu $ by an amount $\Delta \protect\mu $ (broken lines), a first
order theory would provide a correction which diverges when $\protect\mu $
tends to zero.}}{}{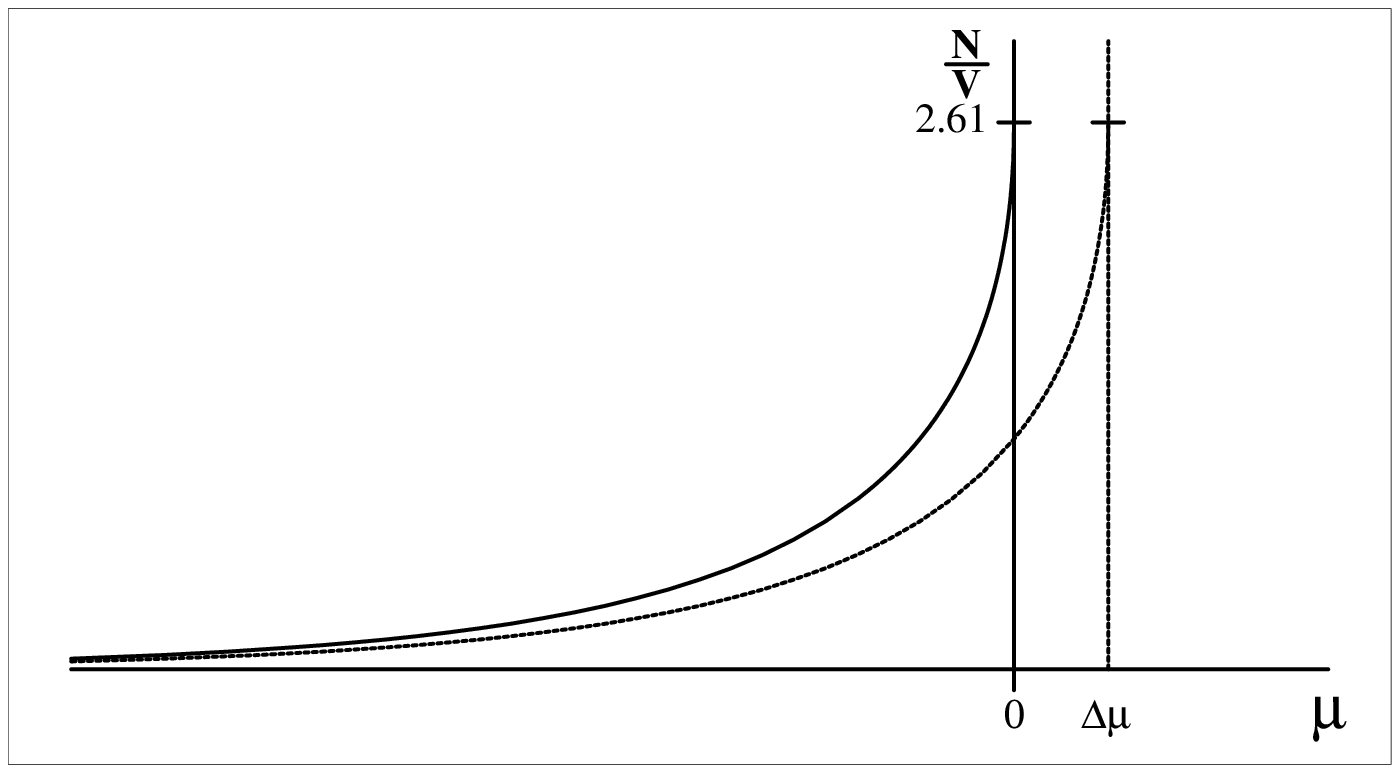}{\special{language "Scientific Word";type
"GRAPHIC";maintain-aspect-ratio TRUE;display "USEDEF";valid_file "F";width
4.3284in;height 2.2589in;depth 0pt;original-width 7.4884in;original-height
3.8951in;cropleft "0";croptop "1";cropright "1";cropbottom "0";filename
'fg1.eps';file-properties "XNPEU";}}

This makes the problem more complicated. To see why, assume for a moment
that the effect of the interactions is just to change this critical value by
a small amount $\Delta \mu $, positive if the interactions are repulsive,
negative if they are attractive. In this case, the behavior of the density
near the critical value of $\mu $ would be given by: 
\begin{equation}
n\simeq c_{1}-c_{2}\sqrt{-\left( \mu +\Delta \mu \right) }  \label{01}
\end{equation}
This function can be expanded in a Taylor series of powers of $\Delta \mu $,
and the result is a series where the term in $(\Delta \mu )^{n}$ is
proportional to $1/\left| \mu \right| ^{n-1/2}$ which, if $\mu $ tends
towards zero (by negative values), diverges more and more strongly when its
order increases; clearly the validity of the expansion breaks down at the
transition point.\ This is a consequence of the non-analicity of the
unperturbed function at the origin (an square root with an infinite
derivative) while, despite of this mathematical problem, the physical nature
of the transition remains completely unaffected by the presence of a small
correction $\Delta \mu $.\ This example illustrates the dangers of using
finite perturbation series near a transition point\footnote{%
It is not necessarily realistic: in fact we will see in this article that
the effects of interactions are more interesting than a simple shift of the
value of the chemical potential.\ It is nevertheless interesting to note
that several of the properties that we will find in our more elaborate study
depend critically on the singularity of the curve giving the density of an
ideal gas.}; this is why we have to go beyond the expressions written in 
\cite{Ursell operators 2} and to use a method where series are summed up to
an infinite order.

\subsection{Operatorial derivative}

The Ursell diagrams that we will use in this article are similar to, but
slightly different from those used in ref.\cite{Ursell operators}: instead
of the grand potential, what will be expanded here is the expression of the
one-particle density operator.\ In any case, the two sorts of diagrams are
closely related, since reduced density operators can be derived from the
grand canonical partition function \cite{Ursell operators 2}.\ But, instead
of first introducing the diagrams and then taking derivatives, it turns out
to be more convenient to start again the calculation from the beginning,
mostly because this avoids introducing weights that, in a second step, will
be cancelled in many cases.

We use the same notation as in \cite{Ursell operators} and \cite{Ursell
operators 2}; the $P_{\alpha }$'s are all the $N!$ permutations of the $N$
particles, which will be expressed as products of permutation cycles; the $%
U_{n}$'s are the Ursell operators.\ The first Ursell operator $U_{1}$ is
defined as function of the one particle hamiltonian $H_{1}$ (kinetic energy
for a gas in a box) as: 
\begin{equation}
U_{1}(1)=\exp \left[ -\beta H_{1}(1)\right]  \label{16}
\end{equation}
while the second operator $U_{2}$ is defined as a function of the
two-particle hamiltonian $H_{2}$ (including interactions between the two
particles) by: 
\begin{equation}
U_{2}(1,2)=\exp \left[ -\beta H_{2}(1,2)\right] -\exp \left[ -\beta H_{1}(1)%
\right] \exp \left[ -\beta H_{1}(2)\right]  \label{16bis}
\end{equation}
(similar expressions can be written for the higher rank Ursell operators).\
Equation (9) of ref. \cite{Ursell operators} provides the following
expression of the $N$ particle partition function: 
\begin{equation}
Z_{N}=\frac{1}{N!}\sum_{\left\{ P_{\alpha }\right\} }\sum_{\left\{ U\right\}
}\prod_{clusters}\,\Gamma _{cluster}(i,j,k...)  \label{1}
\end{equation}
where each of the $\sum $ in equation (\ref{1}) is actually a simplified
notation for two different summations; the sum over $\left\{ U\right\} $ is
meant to contain a sum over all possible ways to write products of $U_{n}$'s
containing altogether $N$ particles\footnote{%
This summation corresponds to the sum over the $m_{l}^{^{\prime }}$'s in the
notation of \cite{Ursell operators}.}, as well as another sum over all
possible non-equivalent ways to put numbered particles into these $U$'s -
see equation (4) of \cite{Ursell operators}; similarly, the sum over $%
\left\{ P_{\alpha }\right\} $ is meant to contain all possible ways to write
products of cycles for $N$ particles\footnote{%
This summation corresponds to the sum over the $m_{l}$'s in equation (7) of 
\cite{Ursell operators}.} as well as all non-equivalent ways to distribute
numbered particles in them - see \S 2 of \cite{Ursell operators} for more
details. The $\Gamma (i,j,k...)$ are the ``explicit'' expressions of the
clusters, in fact traces involving numbered particles $(i,j,k...)$ which are
grouped together into the same cluster by either Ursell operators or
permutation cycles.

To obtain the one-particle density operator, it is convenient to set (for $%
n\geq 2$): 
\begin{equation}
U_{n}(1,2,..n)=\overline{U}_{n}(1,2,..n)\times U_{1}(1)\times U_{1}(2)\times
...U_{1}(n)  \label{2}
\end{equation}
where all the $\overline{U}_{n}$ will be kept constant, while the operators $%
U_{1}$'s will be varied according to: 
\begin{equation}
dU_{1}=dx\times U_{1}\mid \varphi ><\theta \mid  \label{3}
\end{equation}
Here the kets $\mid \varphi >$ and $\mid \theta >$ are any kets in the
one-particle state space. As shown in \cite{Ursell operators 2}, the
one-particle density operator $\rho _{1}^{(N)}$ in the canonical ensemble is
then given by: 
\begin{equation}
<\theta \mid \rho _{1}^{(N)}\mid \varphi >=\frac{1}{Z_{N}}\,\,\frac{d}{dx}%
Z_{N}  \label{4}
\end{equation}
In each term of the double sum in (\ref{1}), one has to take the derivatives
of all $\Gamma $'s with respect to $x$ in succession, which amounts to
taking the derivative with respect to the $U_{1}$ operator of every numbered
particle. The diagram which contains the particle in question then becomes
an operator $\widehat{\Gamma }_{cluster}(i,j,k)$, so that we obtain the
expression: 
\begin{equation}
\rho _{1}^{(N)}=\frac{1}{N!Z_{N}}\sum_{i=1}^{N}\sum_{\left\{ P_{\alpha
}\right\} }\sum_{\left\{ U\right\} }\widehat{\Gamma }_{cluster}(i,j,k)%
\prod_{rest}\Gamma \,_{cluster}(m,p,q...)  \label{4bis}
\end{equation}
where $\prod_{rest}$ symbolizes the product over all remaining $\Gamma
\,_{cluster}$'s which have not be modified by the derivative; sometimes, we
will call this expression ``the triple sum''.

.

\subsection{An example\label{example}}

\bigskip As an example, let us for instance take the following cluster,
corresponding to the diagram shown in figure 2: 
\begin{equation}
\Gamma (1,2,..8)=Tr_{1,2,...8}\left\{ \overline{U}%
_{2}(1,7)U_{1}(1)U_{1}(2).....U_{1}(7)U_{1}(8)C_{6}(1,..6)C_{2}(7,8)\right\}
\label{5}
\end{equation}
where the $C$'s are the permutation cycles - $C_{6}(1,..6)$ is the operator
which creates a circular permutation of particles $1$, $2$, ...$6$ while $%
C_{2}(7,8)$ is merely the exchange operator for particles $7$ and $8$.\
Assume for instance that we take the derivative with respect of particle $%
i=4 $; we then have to evaluate the following expression: 
\begin{equation}
Tr_{1,2,...8}\left\{ \overline{U}_{2}(1,7)U_{1}(1)U_{1}(2)...\widetilde{U}%
_{1}(4)U_{1}(5)..U_{1}(7)U_{1}(8)C_{6}(1,..6)C_{2}(7,8)\right\}  \label{5bis}
\end{equation}
with the notation: 
\begin{equation}
\widetilde{U}_{1}=U_{1}\mid \varphi ><\theta \mid  \label{5ter}
\end{equation}

\FRAME{dtbpFU}{2.3938in}{1.4339in}{0pt}{\Qcb{Fig. 2: An example of an U-C
term in the expansion of the partition function $Z$.}}{\Qlb{fg2}}{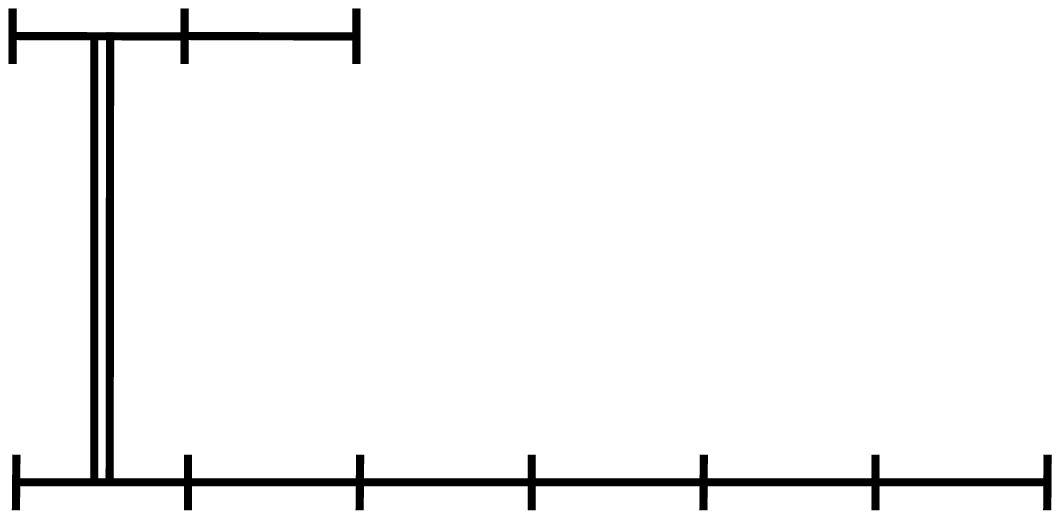}{%
\special{language "Scientific Word";type "GRAPHIC";maintain-aspect-ratio
TRUE;display "USEDEF";valid_file "F";width 2.3938in;height 1.4339in;depth
0pt;original-width 5.7424in;original-height 3.4264in;cropleft "0";croptop
"1";cropright "1";cropbottom "0";filename 'fg2.eps';file-properties "XNPEU";}%
}\ Expression (\ref{5bis}) is nothing but the matrix element: 
\begin{equation}
<\theta \mid \widehat{\Gamma }_{cluster}(1,2,..8)\mid \varphi >
\label{5-ter-2}
\end{equation}
which can be obtained calculated by the same method of calculation as in 
\cite{Ursell operators}, with the introduction of a sufficient number of
closure relations $\sum_{n}\mid u_{n}><u_{n}\mid $ in one-particle spaces.\
In this way, one gets the expression: 
\begin{equation}
\begin{array}{r}
\sum_{n_{1},.....,n_{8}}<1:u_{n_{1}}\mid <2:u_{n_{2}}\mid
.......<8:u_{n_{8}}\mid \overline{U}_{2}(1,7)U_{1}(1)U_{1}(2)\times .... \\ 
...\widetilde{U}_{1}(4)\times U_{1}(5)..U_{1}(7)U_{1}(8)\mid
1:u_{n_{2}}>\mid 2:u_{n_{3}}> \\ 
\times ...\mid 5:u_{n_{6}}>\mid 6:u_{n_{1}}>\mid 7:u_{n_{8}}>\mid
8:u_{n_{7}}>
\end{array}
\label{5-4}
\end{equation}
Considering first particles $1$, we see that the circular permutation of
indices in the kets at then end of the expression, together with the
summation over indices $n_{2}$, $n_{3}$,...$n_{6}$, introduce the product of
operators: 
\begin{equation}
\overline{U}_{2}(1,7)\left[ U_{1}(1)\right] ^{3}\widetilde{U}_{1}(1)\left[
U_{1}(1)\right] ^{2}  \label{5-5}
\end{equation}
while the summation over $n_{1}$ introduces a trace over particle $1$;
similarly, the summation over $n_{8}$ introduces the product of two
additional operators $U_{1}(7)$. Finally we get: 
\begin{equation}
Tr_{1,7}\left\{ \overline{U}_{2}(1,7)\left[ U_{1}(1)\right] ^{3}\widetilde{U}%
_{1}(1)\left[ U_{1}(1)\right] ^{2}\left[ U_{1}(7)\right] ^{2}\right\}
\label{5-6}
\end{equation}
-which is nothing but the matrix element (the dummy index $7$ is renamed
into $2$): 
\begin{equation}
<1:\theta \mid Tr_{2}\left\{ \left[ U_{1}(1)\right] ^{2}\left[ U_{1}(2)%
\right] ^{2}\overline{U}_{2}(1,2)\left[ U_{1}(1)\right] ^{3}U_{1}(1)\right\}
\mid 1:\varphi >  \label{5-7}
\end{equation}
Finally we obtain: 
\begin{equation}
\begin{array}{r}
\widehat{\Gamma }_{cluster}(1,2,..8)=Tr_{2}\left\{ \left[ U_{1}(1)\right]
^{2}\overline{U}_{2}(1,2)\left[ U_{1}(1)\right] ^{4}\left[ U_{1}(2)\right]
^{2}\right\} \\ 
=\left[ U_{1}(1)\right] ^{2}Tr_{2}\left\{ U_{2}(1,2)U_{1}(2)\right\} \left[
U_{1}(1)\right] ^{3}
\end{array}
\label{6}
\end{equation}
This is, of course, an operator - no longer a number as was the initial $%
\Gamma $; but it can also be represented by a diagram, such as that shown in
figure 3.\ In this new diagram, the lowest horizontal line no longer
corresponds to a trace but to a product of operators $U_{1}(1)$ - as many as
there are segments in this line (two in this particular case) - interrupted
at some point by a $U_{2}(1,2)$, and then followed by another horizontal
line symbolizing again the product of $U_{1}$'s (three in this case). On the
other hand, the upper horizontal line still corresponds to a trace over
particle $2$; it also contains operators $U_{1}(2)$, but here they all
remain after the operator $U_{2}(1,2)$.\FRAME{dtbpFU}{2.1897in}{1.1701in}{0pt%
}{\Qcb{Fig. 3: An example of an U-C term in the expansion of the
one-particle density-operator $\protect\rho _{1}$.}}{\Qlb{fg3}}{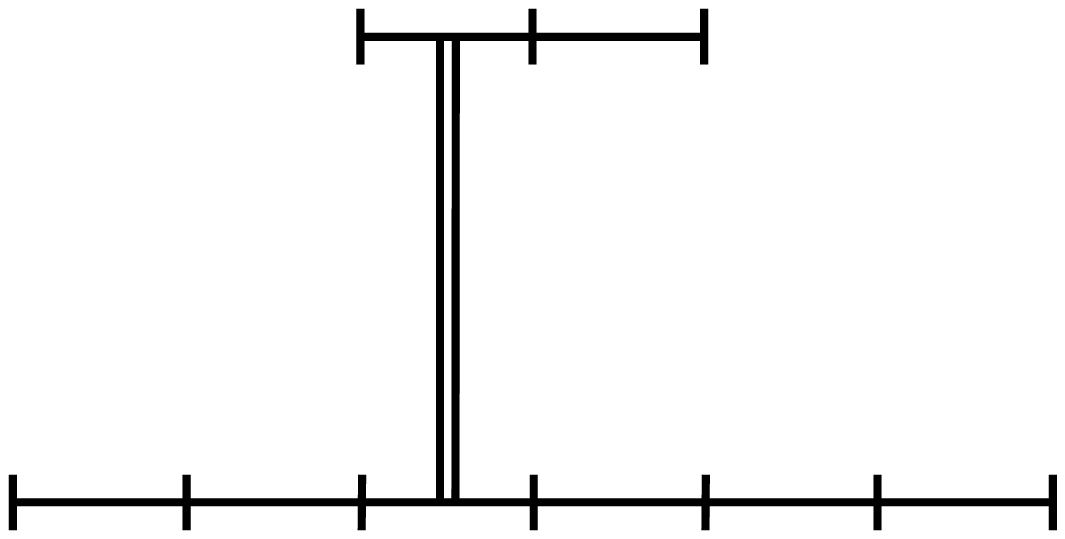}{%
\special{language "Scientific Word";type "GRAPHIC";maintain-aspect-ratio
TRUE;display "USEDEF";valid_file "F";width 2.1897in;height 1.1701in;depth
0pt;original-width 5.4769in;original-height 2.9127in;cropleft "0";croptop
"1";cropright "1";cropbottom "0";filename 'fg3.eps';file-properties "XNPEU";}%
}

\subsection{Diagrams; weights}

More generally, all operators $\widehat{\Gamma }$ in the triple sum (\ref
{4bis}) can be represented by diagrams where the lowest horizontal line
represents a product of operators, beginning by either a chain of $U_{1}$'s,
or a single $U_{1}$, or actually any $U_{n}$; this is the main difference
with diagrams contained in $Z$ (or $\log Z$), where the initial operator was
always that of highest rank. All the other horizontal lines correspond to
traces, and begin necessarily with an operator or rank at least $n=2$; in
fact, these lines behave exactly as those in $\log Z$. In other words the
new diagrams are, so to say, obtained by ``cutting'', or ``opening'' the old
diagrams at some arbitrary point, taken in some exchange cycle\footnote{%
This cycle may of course be of length one, corresponding to no exchange at
all.}, which then no longer corresponds to a trace but to a product of
operators - see the two examples shown in figure 4.

As in \cite{Ursell operators}, we need a rule to decide precisely how the
diagrams should be drawn; this is necessary in order to avoid ambiguities
and double counting.\ Fortunately, the fact that the derivative has
``tagged'' one particle - adding one summation to the two which already
exist in (\ref{1}) - makes the problem much simpler.\ The tagged particle
determines the first operator of the lowest exchange cycle; it provides a
well defined starting point, a ``root'' from which one can propagate
horizontally along exchange cycles and vertically along $U_{2}$'s or
operators of higher rank. Actually, when only $U_{2}$ operators are present
in the diagram, no ambiguity ever occurs: the simple propagation from the
root into the branches is sufficient to assign a well defined structure to
each term - see for instance the first example of figure 4.\ Only when
operators $U_{n}$ of rank $n$ equal to $3$ (or more) occur, as in the second
example of this figure, can some ambiguity occur: which is cycle is in the
middle, which one in the upper position? By similarity with what was done in 
\cite{Ursell operators}, we take the following convention: beyond the first
particle, which is determined by propagation along a previous cycle, all the
other particles in $U_{n}$'s of rank equal to $3$ (or more) appear in order
of increasing numbering; in other words, in the vertical lines, particles
appear with an upward increasing numbering, except of course for the lowest
particle which is either the tagged particle, or is determined from it by
propagation along previous cycles and $U_{n}$'s. With this rule combined
with those of \cite{Ursell operators}, each contribution to $\rho _{1}^{(N)}$
appearing in the triple sum (\ref{4bis}) corresponds to a perfectly well
defined series of diagrams, starting with an operator $\widehat{\Gamma }%
_{\rho }$, and followed by a product of numbers $\Gamma _{diag.}$.

\FRAME{dtbpFU}{3.0692in}{1.4209in}{0pt}{\Qcb{Fig. 4: Two examples of
diagrams occurring in the expansion of $\protect\rho _{1}$; the diagram on
the left has a weight 1, but that on the right has a weight $1/2$.}}{\Qlb{fg4%
}}{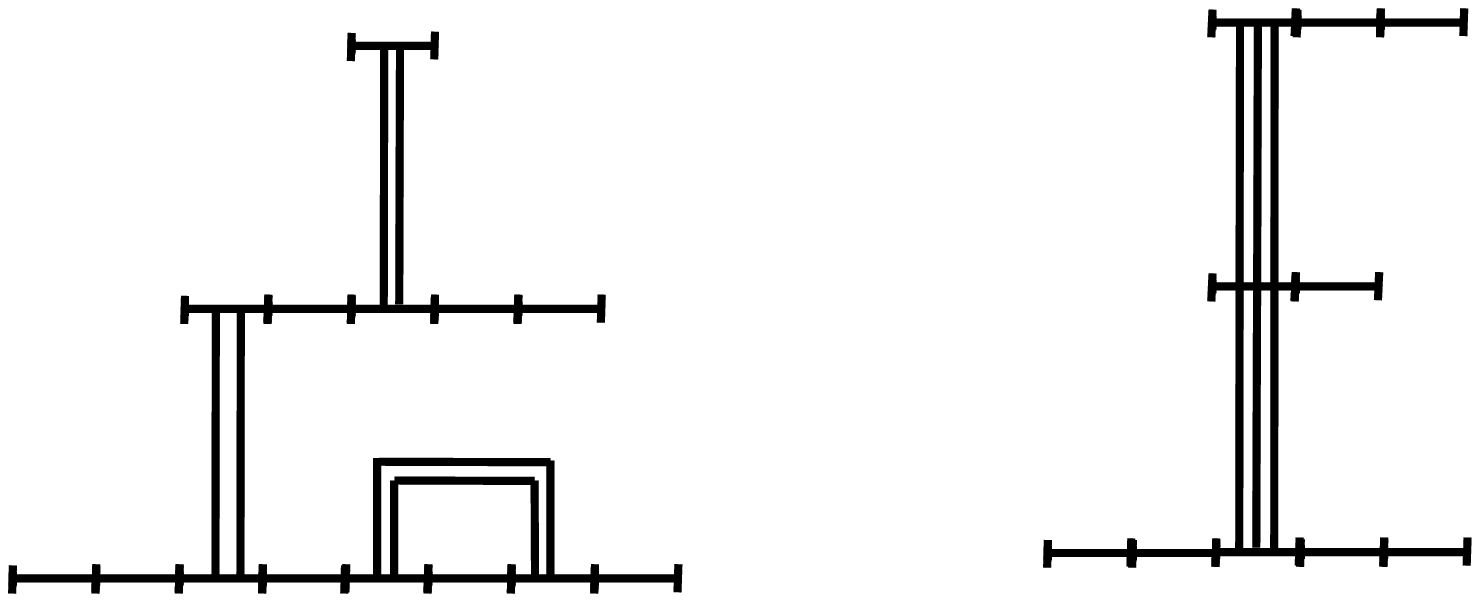}{\special{language "Scientific Word";type
"GRAPHIC";maintain-aspect-ratio TRUE;display "USEDEF";valid_file "F";width
3.0692in;height 1.4209in;depth 0pt;original-width 7.1278in;original-height
3.2828in;cropleft "0";croptop "1";cropright "1";cropbottom "0";filename
'fg4.eps';file-properties "XNPEU";}}

Conversely, if we did not specify any numbering in the diagrams, it is clear
that the same series of diagrams corresponds to many different terms in the
triple sum (\ref{4bis})\footnote{%
In fact, a given $\widehat{\Gamma }_{\rho }$ may even originate from several
different $\Gamma _{diag.}$'s; but this is no longer true as soon as it
contains numbered sites (the lowest of all numbers in the $U_{n}$ of highest
rank will determine the root of the $\widehat{\Gamma }_{\rho }$).}; in order
to get a one-to-one correspondence, we have to specify particle numbers at
each location inside the diagrams, in other words to consider ``numbered
diagrams'' where each ``site'' in the diagrams gets a number. This can be
done precisely by the rules mentioned in the preceding paragraph: starting
from the tagged particle in $\widehat{\Gamma }_{\rho }$, one adds
horizontally the particles it exchanges with (in the order of the cyclic
permutation), and then progresses vertically along higher order rank $U_{n}$%
's: this provides a unique distribution of all relevant particles - as for
the $\Gamma _{diag.}$'s, they are of course treated exactly as in \cite
{Ursell operators}.\ We can then write: 
\begin{equation}
\rho _{1}^{(N)}=\frac{1}{N!Z_{N}}\sum_{diag.}\sum_{distrib.}\widehat{\Gamma }%
_{\rho }(\;.\;,\;.\;,\;.\;)\prod_{rest}\Gamma _{diag.}(\;.\;,\;.\;,\;.\;)
\label{6bis}
\end{equation}
where $\sum_{diag.}$ stands for all the possible ways to write series of
diagrams, one $\Gamma _{\rho }$ and one arbitrary number of $\Gamma _{diag.}$%
's, for a total number of particles equal to $N$ ; in addition, $%
\sum_{distrib.}$ introduces a summation over all correct ways to distribute
numbered particles into the series.\ One should keep in mind that any random
distribution of numberings in a diagram $\widehat{\Gamma }_{\rho }$ is not
necessarily acceptable: in general; in fact, only a proportion $g_{\rho }$
is correct (similarly, only a proportion $f_{diag.}$ is acceptable for
diagrams $\Gamma _{diag.}$, as discussed in \cite{Ursell operators}).\ We
call this proportion the weight of the diagram; for instance, if a diagram
contains one $U_{3}$, the value of $g_{\rho }$ is $1/2$; if it contains $p$
operators $U_{3}$'s its value is $1/2^{p}$; if it contains $q$ operators $%
U_{4}$, its value is $1/6^{q}$, etc...

\subsection{Counting diagrams}

It is now possible to get rid of particle numbering which, of course, does
not affect the contribution to $\rho _{1}$ of any particular term. Let us
choose one given (non-numbered) diagram $\Gamma _{\rho }$, call $n_{\rho }$
the number of particles it contains, and calculate the total coefficient
that it gets from the summations of (\ref{6bis}): 
\begin{equation}
\frac{1}{N!}\sum_{distrib.}\widehat{\Gamma }_{\rho
}(.,.,.)\sum_{diag.}\prod_{rest}\Gamma _{diag.}(.,.,.,)  \label{7bis}
\end{equation}
This can be done in two steps: first calculate the contribution of the term $%
\Gamma _{\rho }(i,j,k)$ when the particles $(i,j,k)$ are a given
sub-ensemble of the $N$ particles; then sum over all the ways to select this
sub-ensemble.\ The first step can be made by remarking that, for the $%
N-n_{\rho }$ remaining particles, the $\sum_{distrib.}$ can be moved to the
right of $\Gamma _{\rho }$ , which introduces the same expression as in \cite
{Ursell operators}; therefore the summation of the products of $\Gamma
_{diag.}$'s merely reconstructs the partition function $Z_{N-n_{\rho }\text{ 
}}$of $N-n_{\rho }$ particles, multiplied by $(N-n_{\rho })!$. As for the $%
n_{\rho }$ particles, one has to take into account the number of correct
distributions of the $n_{\rho }$ particles in the first diagram, which is $%
(n_{\rho })!g_{\rho }$.\ For the second step, we first have to multiply the
result by the number of ways to distribute $n_{\rho }$ particles among $N$,
which is: 
\begin{equation}
\frac{N!}{(N-n_{\rho })!(n_{\rho })!}  \label{7ter}
\end{equation}
Taking all these factors into account, we obtain the following result: 
\begin{equation}
\frac{1}{N!Z_{N}}\frac{N!}{(N-n_{\rho })!(n_{\rho })!}\,(n_{\rho
})!\,\,g_{\rho }\times (N-n_{\rho })!\,\,Z_{N-n_{\rho }}\times \widehat{%
\Gamma }_{\rho }  \label{9}
\end{equation}
so that our final expression for the canonical ensemble is: 
\begin{equation}
\rho _{1}^{(N)}=\sum_{\rho \,\,\,diagrams}\frac{Z_{N-n_{\rho }}}{Z_{N}}%
\,\,g_{\rho }\,\,\widehat{\Gamma }_{\rho }  \label{10}
\end{equation}

At this point, it becomes convenient to introduce the grand canonical
ensemble and its partition function: 
\begin{equation}
Z_{g.c.}=\sum_{N}z^{N}Z_{N}  \label{11}
\end{equation}
where $z=e^{\beta \mu }$ is, with usual notation, the fugacity; the
corresponding one-particle density operator $\rho _{1}$ (to simplify the
notation, we now give up the index $g.c.$): 
\begin{equation}
<\theta \mid \rho _{1}\mid \varphi >=\,\,\frac{1}{Z_{g.c.}}\frac{d}{dx}%
Z_{g.c.}=\frac{1}{Z_{g.c.}}\sum_{N}z^{N}Z_{N}<\theta \mid \rho
_{I}^{(N)}\mid \varphi >  \label{12}
\end{equation}
We then get:

\begin{equation}
\begin{array}{ll}
\displaystyle\rho _{1} & \displaystyle=\frac{1}{Z_{g.c.}}\sum_{N}z^{N}Z_{N}%
\sum_{\rho \,\,\,diagrams}g_{\rho }\frac{Z_{N-n_{\rho }}}{Z_{N}}\widehat{%
\Gamma }_{\rho } \\ 
& \displaystyle=\frac{1}{Z_{g.c.}}\sum_{\rho \,\,\,diagrams}z^{n_{\rho
}}\,g_{\rho }\,\widehat{\Gamma }_{\rho }\sum_{N-n_{\rho }}\,z^{N-n_{\rho
}}\,Z_{N-n_{\rho }}
\end{array}
\label{13}
\end{equation}
But the second summation reconstructs is just another expression of the
grand canonical partition function, so that we finally obtain: 
\begin{equation}
\displaystyle\rho _{1}=\sum_{\rho \,\,\,diagrams}z^{n_{\rho }}\,g_{\rho }\,%
\widehat{\Gamma }_{\rho }  \label{14}
\end{equation}
This expression shows that the one-particle density operator is merely the
sum of all diagrams $\widehat{\Gamma }_{\rho }$, with coefficients which are
the product of the weight $g_{\rho }$ by the fugacity raised to a power
which is the number of particles contained in the diagram.

The simplest case is, of course, the ideal gas, for which the only diagrams
are horizontal lines; we immediately get: 
\begin{equation}
\displaystyle\rho _{1}=\sum_{n=1}^{\infty }z^{n}\left[ U_{1}\right]
^{n}=f_{1}  \label{14bis}
\end{equation}
where $f_{1}$ is defined as usual by: 
\begin{equation}
f_{1}=\frac{zU_{1}}{1-zU_{1}}  \label{15}
\end{equation}
Actually, even for interacting gases, this operator plays a role in the
diagrams, or more precisely the sum: 
\begin{equation}
\sum_{n=0}^{\infty }z^{n}\left[ U_{1}\right] ^{n}=1+f_{1}  \label{15bis}
\end{equation}
which can be connected for instance after any operator $U_{2}$. We take the
same convention as in \cite{Ursell operators}: dashed horizontal lines
symbolize this operator; the operator $f_{1}$ itself is then represented by
a dashed line connected to a one-case horizontal line ($f_{1}=zU_{1}+\left[
1+f_{1}\right] $). The first terms in the Ursell expansion of $\rho _{I}$
are shown in figure 5; those explicitly shown in the first and second line
correspond to the ``generalized Beth-Uhlenbeck approximation discussed in
this reference, which in terms of $\rho _{1}$ would lead to the expression: 
\begin{equation}
\rho _{1}=f_{1}+2z^{2}\,\left[ 1+f_{1}(1)\right] Tr_{2}\left\{ U_{2}^{S}(1,2)%
\left[ 1+f_{1}(2)\right] \right\} \left[ 1+f_{1}(1)\right]  \label{15ter}
\end{equation}
A diagram where the same $U_{2}$ re-connects to the same cycle corresponds
to an exchange term, where $U_{2}$ is replaced by its product by the
exchange operator $P_{ex.}$, as discussed in \cite{Ursell operators}; for
instance, in the second line of fig. 5, the second diagram is merely the
exchange term of the first.

\FRAME{dtbpFU}{3.4108in}{2.1862in}{0pt}{\Qcb{Fig. 5: The lowest order terms
in the U-C expansion of the one-particle density-operator $\protect\rho $.}}{%
\Qlb{fg5}}{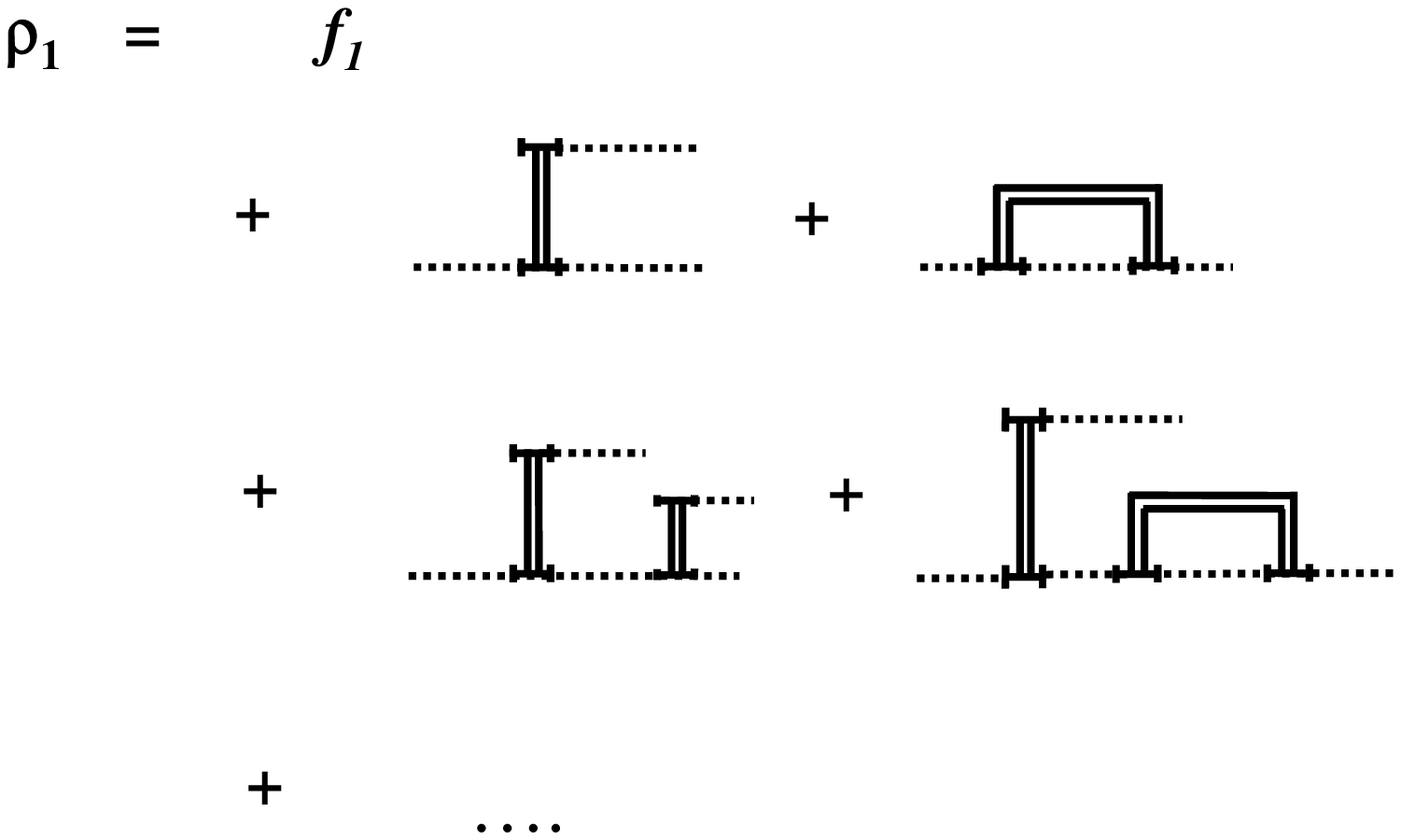}{\special{language "Scientific Word";type
"GRAPHIC";maintain-aspect-ratio TRUE;display "USEDEF";valid_file "F";width
3.4108in;height 2.1862in;depth 0pt;original-width 7.1157in;original-height
4.5515in;cropleft "0";croptop "1";cropright "1";cropbottom "0";filename
'fg5.eps';file-properties "XNPEU";}}

\section{Integral equations\label{integral}}

The largest eigenvalue of $U_{1}$ corresponds to the ground state of the
one-particle hamiltonian $H_{1}$.\ When the chemical potential reaches the
energy of this state (zero for a gas in a box with periodic boundary
conditions), the action of each operator $zU_{1}$ leaves the ground state
unchanged; then, the series of this operator which is contained in $f_{1}$
(sum of all of its powers) diverges.\ This is not surprising since, as
already mentioned, Bose-Einstein condensation corresponds to a situation
where the size of typical exchange cycles is no longer limited to a few
units, but increases to infinity.\ For an interacting system, the phenomenon
is not limited to chains of $U_{1}$'s only: when very long exchange cycles
become important, it is expected that typical diagrams will also contain an
arbitrary large number of operators $U_{2}$ (or higher rank Ursell
operators) connecting these cycles.\ In practice, this means that it is not
possible to limit the calculations to a few diagrams, those shown explicitly
in figure 5 for instance; a summation over large categories of diagrams is
indispensable.

This is what is done in the two sections below, first within a simplified
approach which is useful as a preliminary step (it will lead us to a
mean-field like theory), second within a more elaborate treatment.

\subsection{A first approach: Dyson-Ursell equation\label{Dyson}}

For simplicity, we limit ourselves for the moment to terms which contain
Ursell operators of rank 1 and 2, while a generalization to higher ranks is
possible in a similar way.\ The symmetrized version of $U_{2}$ is: 
\begin{equation}
U_{2}^{S}=\frac{1}{2}\left[ 1+P_{ex.}\right] U_{2}=\frac{1}{2}U_{2}\left[
1+P_{ex.}\right]  \label{17}
\end{equation}
where $P_{ex.}$ is the exchange operator of the two particles contained in $%
U_{2}$ and $\eta $ is equal to $+1$ for bosons (actually the only case we
study here) and $-1$ for fermions; the factor 1/2 in front of this
definition is introduced for the first part of the formula to be a projector
(if necessary, one can put projectors on each side of $U_{2}$ without
changing the result).

Let us now write the following implicit equation: 
\begin{equation}
\rho _{1}(1)=f_{1}(1)+2z^{2}\,\left[ 1+f_{1}(1)\right] Tr_{2}\left\{
U_{2}^{S}(1,2)\left[ 1+\rho _{1}(1)\right] \left[ 1+\rho _{1}(2)\right]
\right\}  \label{18}
\end{equation}
and investigate which diagrams are contained in it. The zero order term in $%
U_{2}$ is merely $f_{1}$; the first order terms in $U_{2}$ are simply the
second and the third diagram of figure 5 - in \cite{Ursell operators} we
showed how terms introduced by the exchange operator correspond to diagrams
where the same $U_{2}$ touches twice the same horizontal exchange cycle.
More generally, it is easy to see that the term $\rho _{1}^{(n)}$of order $n$
in $U_{2}$ is obtained as a function of terms of lower order by the relation%
\footnote{%
The equation is written for the case where $n>1$; if $n=1$, the curly
bracket of (\ref{19}) is simply equal to $\left\{ U_{2}^{S}\left[ 1+f_{1}(1)%
\right] \left[ 1+f_{1}(2)\right] \right\} $. A more precise discussion is
given in Appendix A}: 
\begin{equation}
\begin{array}{ll}
\rho _{1}^{(n)}= & 2z^{2}\left[ 1+f_{1}(1)\right] Tr_{2}\left\{ U_{2}^{S}%
\left[ 1+f_{1}(1)\right] \rho _{1}^{(n-1)}(2)\right. \\ 
& \left. +U_{2}^{S}\rho _{1}^{(1)}(1)\rho
_{1}^{(n-2)}(2)+.....+U_{2}^{S}\rho _{1}^{(n-1)}(1)\left[ 1+f_{1}(2)\right]
\right\}
\end{array}
\label{19}
\end{equation}
In the right hand side, let us first consider the direct term, where $%
U_{2}^{S}$ is replaced by $U_{2}$: diagrammatically, in the two horizontal
lines which follow $U_{2}$, one gets in succession all possible combinations
of lower order ``branches''.\ For instance, two second order terms ($n=2$)
are shown in the third line of figure 5; two others should follow with an
additional $U_{2}$ connected in two different ways to the upper cycle.\ As
for the exchange term, it gives a diagram where the first $U_{2}$ reconnects
to the initial exchange cycle, so that this time it is the line between the
two contact points and the line after the second point which play the role
of the two lines after the $U_{2}$; nevertheless the situation remains
essentially the same with a different topology (taking again the example of
second order terms, one would get 4 more diagrams where the first $U_{2}$ is
folded over the same initial cycle). We then see that, when the iteration is
continued to infinity, it provides once, and once only, all possible
``branched'' diagrams with an arbitrary number of $U_{2}$'s: indeed,
equation (\ref{19}) provides a convenient way to perform a summation
containing an infinite number of diagrams and cycles with infinite length.\
Nevertheless one should keep in mind that this is not exact either: more
complicated diagrams, which are not simply ``branched'', are still not
included, for instance those containing loops such as those of figure 6.%
\FRAME{dtbpFU}{3.0943in}{1.3975in}{0pt}{\Qcb{Fig.\ 6: Diagrams containing
loops which are not included in the Ursell-Dyson approximation, but will be
in the more elaborate integral equation of \S\ \ref{more}; the second
diagram corresponds to the exchange term of the first.}}{\Qlb{fg6}}{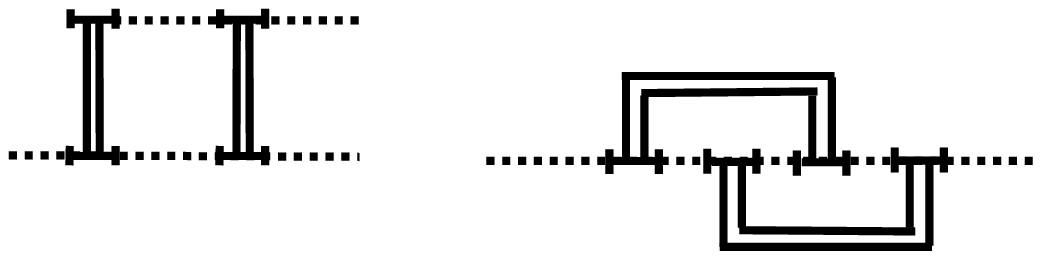}{%
\special{language "Scientific Word";type "GRAPHIC";maintain-aspect-ratio
TRUE;display "USEDEF";valid_file "F";width 3.0943in;height 1.3975in;depth
0pt;original-width 5.4328in;original-height 2.4388in;cropleft "0";croptop
"1";cropright "1";cropbottom "0";filename 'fg6.eps';file-properties "XNPEU";}%
}

Remark: equation (\ref{18}) is a simple generalization of (\ref{15ter}), but
one could think of other simple iteration equations\footnote{%
Another remark is that the $1$'s are indispensable in (\ref{18}); otherwise
diagrams where $U_{2}$ is not directly followed with $U_{1}^{\prime }$'s
would be excluded.}, for instance: 
\begin{equation}
\rho _{1}(1)=f_{1}(1)+2z^{2}\,\left[ 1+\rho _{1}(1)\right] Tr_{2}\left\{
U_{2}^{S,A}(1,2)\left[ 1+\rho _{1}(1)\right] \left[ 1+\rho _{1}(2)\right]
\right\}  \label{20}
\end{equation}
(the only difference with (\ref{18}) is that one $f_{1}$ has been replaced
by one $\rho _{1}$).\ Nevertheless, one can easily convince oneself that
this equation is not satisfactory since it just introduces redundancy,
actually already at second order in $U_{2}$.

\subsection{A more elaborate integral equation\label{more}}

The approximation made in the preceding section can be completed by adding
terms where $U_{2}$ operators connect twice the same pair of cycles so that
they introduce loops. In fact, to include them, it is sufficient to add the
following term to the right hand side of (\ref{18}): 
\begin{equation}
\begin{array}{r}
z^{4}\left[ 1+f_{1}(1)\right] Tr_{2}\left\{ \left[ U_{2}(1,2)\left[ 1+\rho
_{1}(1)\right] \left[ 1+\rho _{1}(2)\right] ^{\stackrel{\stackrel{}{}}{}}%
\right] ^{2}\left[ 1+P_{ex.}\right] \right\} \\ 
=\,\,\,\,2z^{4}\left[ 1+f_{1}(1)\right] Tr_{2}\left\{ \left[ U_{2}^{S}(1,2)%
\left[ 1+\rho _{1}(1)\right] \left[ 1+\rho _{1}(2)\right] ^{\stackrel{%
\stackrel{}{}}{}}\right] ^{2}\right\}
\end{array}
\label{21}
\end{equation}
(the second line is easily obtained by introducing the symmetrizer $\left[
1+P_{ex.}\right] /2$, which is a projector and can therefore be raised to
any power, and using the commutation between this operator and $U_{2}$).
When this is done, diagrams with an arbitrary number of pairs of $U_{2}$'s
touching twice the same two cycles are included, as illustrated by the
example shown in figure 7. Note that, since the right hand side of equation (%
\ref{19}) now contains the sum of two terms, any combination remains
possible: some cycles may be connected only once by a $U_{2}$, others
twice.\ We note in passing that, in these new diagrams, every $U_{2}$ does
not necessarily introduce a trace over a new particle, as opposed to the
situation in branched diagrams.

There is no special difficulty in including now terms with an arbitrary
number $q$ of operators $U_{2}$'s forming a ladder between the same two
cycles; one now has to add the following term to the right hand side of (\ref
{19}): 
\begin{equation}
\begin{array}{r}
z^{2q}\left[ 1+f_{1}(1)\right] Tr_{2}\left\{ \left[ U_{2}(1,2)\left[ 1+\rho
_{1}(1)\right] \left[ 1+\rho _{1}(2)\right] ^{\stackrel{\stackrel{}{}}{}}%
\right] ^{q}\left[ 1+P_{ex.}\right] \right\} \\ 
=2z^{2q}\left[ 1+f_{1}(1)\right] Tr_{2}\left\{ \left[ U_{2}^{S}(1,2)\left[
1+\rho _{1}(1)\right] \left[ 1+\rho _{1}(2)\right] ^{\stackrel{\stackrel{}{}%
}{}}\right] ^{q}\right\}
\end{array}
\label{22}
\end{equation}
The summation of all these contributions, from $q=1$ as in \S\ \ref{Dyson}
up to infinity, is possible and provides the following generalization of (%
\ref{18}): 
\begin{equation}
\rho _{1}(1)=f_{1}(1)+2z^{2}\left[ 1+f_{1}(1)\right] Tr_{2}\left\{ \frac{%
U_{2}^{S}(1,2)\left[ 1+\rho _{1}(1)\right] \left[ 1+\rho _{1}(2)\right] }{%
1-z^{2}U_{2}^{S}(1,2)\left[ 1+\rho _{1}(1)\right] \left[ 1+\rho _{1}(2)%
\right] }\right\}  \label{23}
\end{equation}
At this level of approximation, diagrams similar to that of figure 7 with a
completely arbitrary number of vertical lines connecting the same two cycles
are generated. It is interesting to note that the summation of diagrams
introduces into the trace an operator which can be seen as a two body
generalization of the one particle distribution function of the ideal gas;
one can easily see that more complex terms would introduce three particle
terms, etc..\FRAME{dtbpFU}{3.1142in}{1.8213in}{0pt}{\Qcb{Fig. 7: A typical
diagram containing several loops, which is included in equation (\ref{23}).}%
}{\Qlb{fg7}}{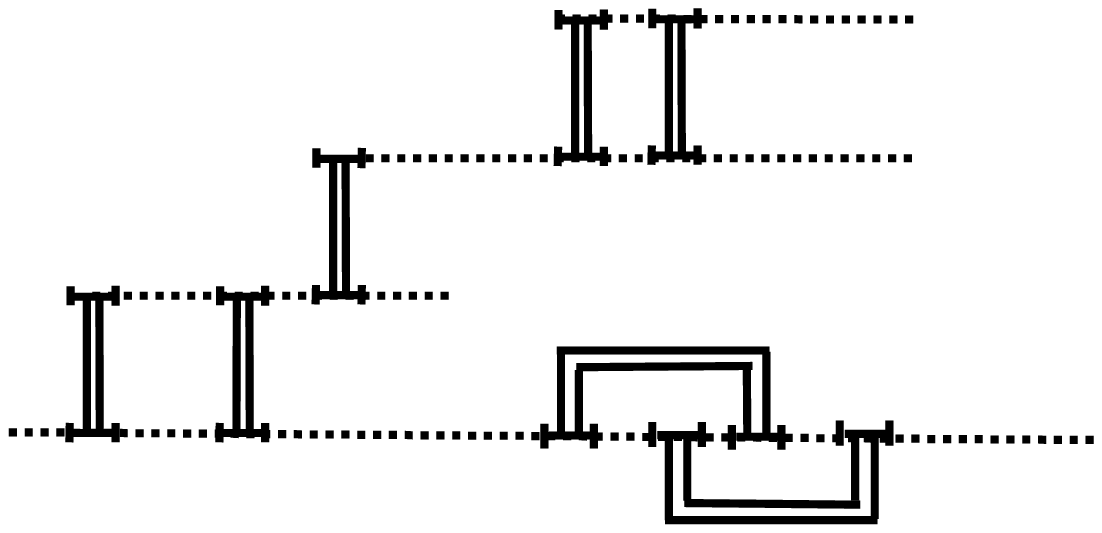}{\special{language "Scientific Word";type
"GRAPHIC";maintain-aspect-ratio TRUE;display "USEDEF";valid_file "F";width
3.1142in;height 1.8213in;depth 0pt;original-width 5.7847in;original-height
3.371in;cropleft "0";croptop "1";cropright "1";cropbottom "0";filename
'fg7.eps';file-properties "XNPEU";}}

\subsection{Discussion}

Clearly, the fact that almost all weights are eliminated from the diagrams
for the density operator (if no other operator than $U_{1}$ and $U_{2}$ is
considered) is the source of great simplification in all calculations;
similar calculations for the expansion of $\log Z$ would have been much more
difficult.\ If necessary, one could readily include in the integral equation
more complicated terms, such as for instance that corresponding to the
diagrams of figure 8, or the second diagram of figure 4, chained an
arbitrary number of times.\ We will see, nevertheless, that in dilute gases
only a limited ensemble of diagrams play a role in the determination of the
critical temperature.\FRAME{dtbpFU}{3.5656in}{1.1441in}{0pt}{\Qcb{Fig 8:
Diagrams that are not included in equation (\ref{23}).}}{\Qlb{fg8}}{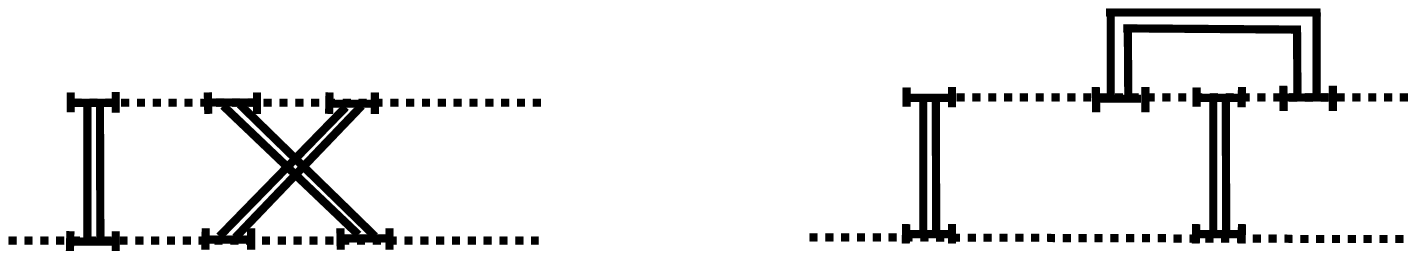}{%
\special{language "Scientific Word";type "GRAPHIC";maintain-aspect-ratio
TRUE;display "USEDEF";valid_file "F";width 3.5656in;height 1.1441in;depth
0pt;original-width 7.0733in;original-height 2.2511in;cropleft "0";croptop
"1";cropright "1";cropbottom "0";filename 'fg8.eps';file-properties "XNPEU";}%
}

\section{A mean-field like approximation\label{mean}}

We first study equation (\ref{18}) and calculate the value of the diagonal
elements $\rho _{\mathbf{k}}$ of the one-particle density operator;
translational invariance in a box with periodic boundary conditions ensures
that the plane waves $\mid \mathbf{k}>$ are the eigenstates of $\rho _{1}$.
These diagonal elements are then: 
\begin{equation}
\rho _{\mathbf{k}}=<\mathbf{k}\mid \rho _{1}\mid \mathbf{k>}  \label{24}
\end{equation}
while, for the ideal gas, they have the value: 
\begin{equation}
f_{\mathbf{k}}=<\mathbf{k}\mid f_{1}\mid \mathbf{k>\,}=\mathbf{\,}\frac{%
\text{e}^{-\beta \widetilde{e}_{k}}}{1-\text{e}^{-\beta \widetilde{e}_{k}}}%
\,\,\mathbf{=\,\,}\frac{1}{\text{e}^{\beta \widetilde{e}_{k}}-1}  \label{25}
\end{equation}
where $f_{1}$ has been defined in (\ref{15}) and $\widetilde{e}_{k}$ is the
difference between the energy and the chemical potential: 
\begin{equation}
\widetilde{e}_{k}=e_{k}-\mu =\frac{\hbar ^{2}k^{2}}{2m}-\mu  \label{26}
\end{equation}
and $m$ the mass of the bosons. We will also replace the value of the
diagonal matrix elements of $U_{2}^{S}$ by their expression as a function of
the (symmetrized) Ursell length $a_{U}^{S}(k)$, which is given in equation
(17) and (4) of \cite{Ursell operators 3}: 
\begin{equation}
z^{2}<\mathbf{k,k}^{^{\prime }}\mid U_{2}^{S}\mid \mathbf{k\mathbf{,k}%
^{^{\prime }}>=}-\frac{\lambda _{T}^{2}}{\mathcal{V}}\,\,\text{e}^{-\beta %
\left[ \widetilde{e}_{k}+\widetilde{e}_{k^{^{\prime }}}\right] }\times
a_{U}^{S}(\left| \mathbf{k}-\mathbf{k}^{^{\prime }}\right| )  \label{26bis}
\end{equation}
where $\lambda _{T}$ is the thermal wavelength: 
\begin{equation}
\lambda _{T}=\frac{h}{\sqrt{2\pi mk_{B}T}}  \label{28}
\end{equation}
(with usual notation) and $\mathcal{V}$ the volume of the container.\ But,
as shown in this reference, the symmetric Ursell length is almost constant
at low energies and equal to twice\footnote{%
This factor two corresponds to the well known statistical increase of the
interaction between two identical bosons.} the scattering length $a$ (if
interactions occur via hard cores, $a$ is merely their diameter).\ We
therefore write: 
\begin{equation}
z^{2}<\mathbf{k,k}^{^{\prime }}\mid U_{2}^{S}\mid \mathbf{k\mathbf{,k}%
^{^{\prime }}>=}-2\frac{\lambda _{T}^{2}}{\mathcal{V}}\,\,\text{e}^{-\beta 
\left[ \widetilde{e}_{k}+\widetilde{e}_{k^{^{\prime }}}\right] }\times a
\label{27}
\end{equation}
which is a very good approximation for low temperature gases (except, of
course, if accidental collision resonances occur at very low energies);
moreover we will assume in all the rest of this article that $a$ is positive
(repulsive interactions).

\subsection{A set of non-linear coupled equations}

Projecting (\ref{18}) over the plane waves provides the relation: 
\begin{equation}
\rho _{\mathbf{k}}=f_{\mathbf{k}}+\sum_{\mathbf{k}^{^{\prime }}}\left[ 1+f_{%
\mathbf{k}}\right] \left[ -\frac{4a\lambda _{T}^{2}}{\mathcal{V}}\right]
\,\,\,\text{e}^{-\beta \left[ \widetilde{e}_{k}+\widetilde{e}_{k^{^{\prime
}}}\right] }\left[ 1+\rho _{\mathbf{k}}\right] \left[ 1+\rho _{\mathbf{k}%
^{^{\prime }}}\right]  \label{29}
\end{equation}
It is convenient to introduce the variables: 
\begin{equation}
X_{\mathbf{k}}=\text{e}^{-\beta \widetilde{e}_{k}}\left[ 1+\rho _{\mathbf{k}}%
\right]  \label{30}
\end{equation}
as well as the dimensionless coupling constant: 
\begin{equation}
\alpha =4\frac{\lambda _{T}^{2}}{\mathcal{V}}a  \label{31}
\end{equation}
Dividing both sides of equation (\ref{29}) by the same factor $\left[ 1+f_{%
\mathbf{k}}\right] $ then provides the simpler equation: 
\begin{equation}
\alpha \sum_{\mathbf{k}^{^{\prime }}}X_{\mathbf{k}^{^{\prime }}}X_{\mathbf{k}%
}-\xi _{k}X_{\mathbf{k}}-1=0  \label{34}
\end{equation}
with the notation: 
\begin{equation}
\xi _{k}=1-\text{e}^{\beta \widetilde{e}_{k}}  \label{33}
\end{equation}
These equations can be obtained by stationarity conditions applied to a
function $\Phi $; see appendix B.

If all variables $X_{\mathbf{k}^{^{\prime }}}$ except $X_{\mathbf{k}}$ are
considered as given, $X_{\mathbf{k}}$ is the solution of a simple quadratic
equation: 
\begin{equation}
\alpha (X_{\mathbf{k}})^{2}-\xi _{k}^{^{\prime }}X_{\mathbf{k}}-1=0
\label{34bis}
\end{equation}
with: 
\begin{equation}
\xi _{k}^{^{\prime }}=\xi _{k}-\Delta \overline{\xi }_{k}  \label{36}
\end{equation}
and\footnote{%
The bar in $\Delta \overline{\xi }$ is used to emphasize the mean-field
character of this correction; in the next section we will include in $\xi
_{k}^{^{\prime }}$ additional terms arising from correlations.}: 
\begin{equation}
\Delta \overline{\xi }_{k}=\alpha \sum_{\mathbf{k}^{^{\prime }}\neq \mathbf{k%
}}X_{\mathbf{k}^{^{\prime }}}  \label{37}
\end{equation}
The solution is straightforward: 
\begin{equation}
X_{\mathbf{k}}=\frac{1}{2\alpha }\left[ \xi _{k}^{^{\prime }}+\sqrt{\left(
\xi _{k}^{^{\prime }}\right) ^{2}+4\alpha }\,\,\right]  \label{35}
\end{equation}
but one should keep in mind that this is an implicit solution: $X_{\mathbf{k}%
}$ depends on all the other $X_{\mathbf{k}^{^{\prime }}}$ through $\Delta 
\overline{\xi }_{k}$ (which accounts for the mean repulsion exerted by all
the other levels), while in turn these variables depend on $X_{\mathbf{k}}$
through their own $\Delta \overline{\xi }_{k^{^{\prime }}}$. The
corresponding system of coupled equations will be solved graphically in the
next section; for the moment, we assume for simplicity that all $X_{\mathbf{k%
}^{^{\prime }}}$'s are kept constant and study the variations of $X_{\mathbf{%
k}}$.

Figure 9 shows this variations as a function of $\xi _{k}^{^{\prime }}$.\ We
can distinguish between two different regimes; first, when $\xi
_{k}^{^{\prime }}$ is negative, one gets from (\ref{35}): 
\begin{equation}
X_{\mathbf{k}}\simeq -\,\,\frac{1}{\xi _{k}^{^{\prime }}}=\frac{1}{\text{e}%
^{\beta \widetilde{e}_{k}}-1+\Delta \overline{\xi }_{k}}  \label{38}
\end{equation}
and: 
\begin{equation}
\rho _{\mathbf{k}}=\text{e}^{\beta \widetilde{e}_{k}}X_{\mathbf{k}}-1=\frac{%
1-\Delta \overline{\xi }_{k}}{\text{e}^{\beta \widetilde{e}_{k}}-1+\Delta 
\overline{\xi }_{k}}  \label{39}
\end{equation}
which is very reminiscent of the equations for the ideal gas; for instance,
if $\beta \widetilde{e}_{k}\ll -1$, the population of the level in question
is given by a Boltzmann exponential. Actually, the effect of the ``mean
repulsion'' $\Delta \overline{\xi }_{k}$ can be expressed in terms of a
positive correction to the chemical potential: 
\begin{equation}
\Delta \overline{\mu }_{k}=-\beta ^{-1}\log \left( 1-\Delta \overline{\xi }%
_{k}\right)  \label{40}
\end{equation}
so that, with this notation, (\ref{39}) becomes: 
\begin{equation}
\rho _{\mathbf{k}}=\frac{1}{\text{e}^{\beta \widetilde{e}_{k}+\Delta 
\overline{\mu }_{k}}-1}=f_{1}(k;\mu -\Delta \overline{\mu }_{k})  \label{42}
\end{equation}
The second regime occurs when $\xi _{k}^{^{\prime }}$ crosses zero and
becomes positive: in this process, the population does not diverge, in
contrast to what would happen for the ideal gas, but reaches a new regime
where its value depends explicitly of the coupling constant: 
\begin{equation}
X_{\mathbf{k}}\simeq \frac{\xi _{k}^{^{\prime }}}{\alpha }  \label{43}
\end{equation}
Actually, since $\alpha $ is inversely proportional to the volume, the
population in question becomes extensive, as expected for the ground state
when it undergoes Bose-Einstein condensation.\FRAME{dtbpFU}{11.2467cm}{%
6.7458cm}{0pt}{\Qcb{Fig. 9: variations of $X_{k}$ as a function of $\protect%
\xi _{k}$; when $\protect\xi _{k}$ is positive, $X_{k}$ reaches a regime
where it becomes proportional to the volume of the system.}}{\Qlb{fg9}}{%
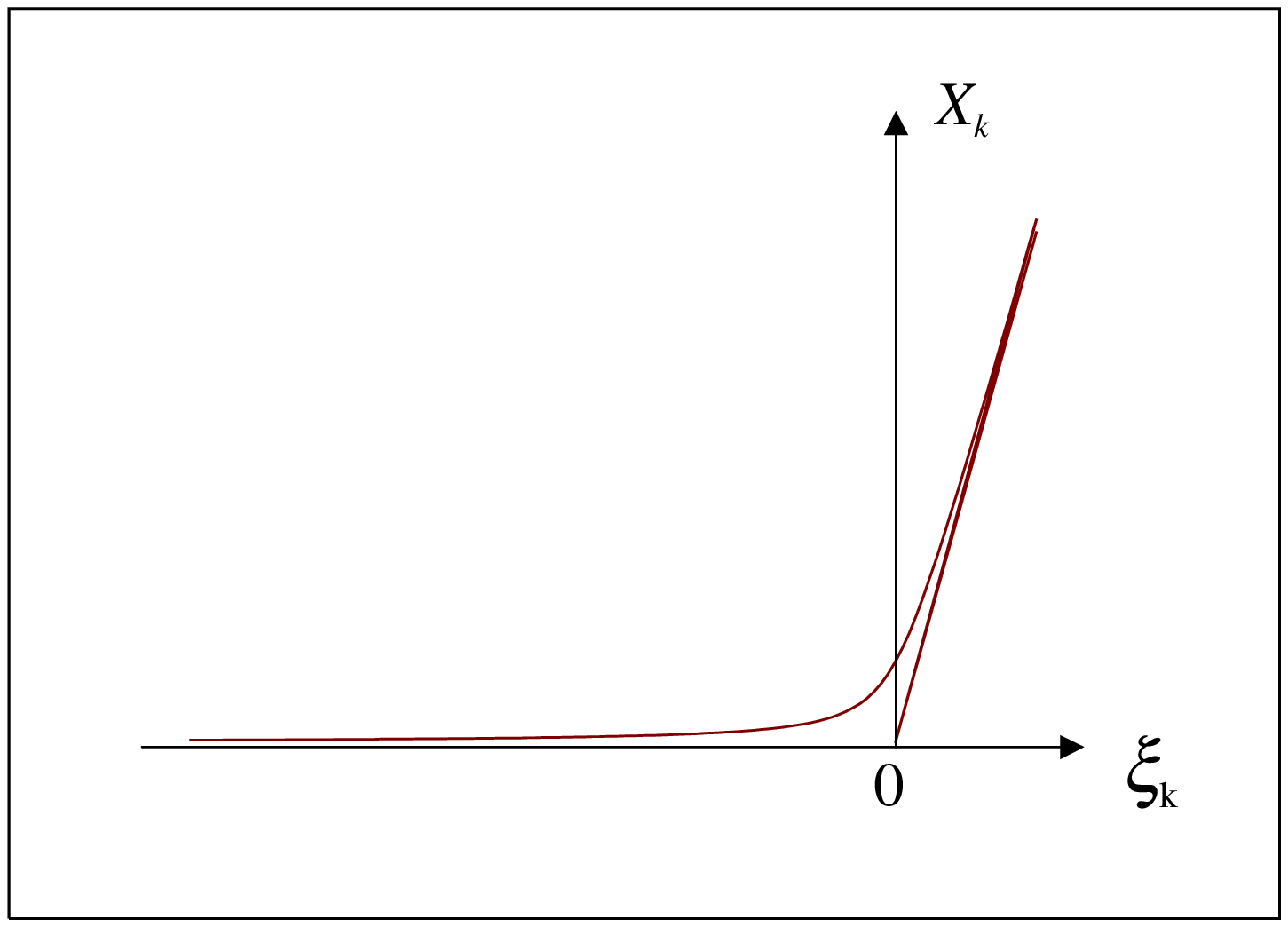}{\special{language "Scientific Word";type
"GRAPHIC";maintain-aspect-ratio TRUE;display "USEDEF";valid_file "F";width
11.2467cm;height 6.7458cm;depth 0pt;original-width 7.2255in;original-height
4.3137in;cropleft "0";croptop "1.0027";cropright "1";cropbottom "0";filename
'fg9.eps';file-properties "XNPEU";}}

Equations (\ref{37}) and (\ref{38}) provide the system of coupled equations
to be solved: each $X_{\mathbf{k}}$ depends on the value of its associated $%
\Delta \overline{\xi }_{k}$, which in turn depends on all the other $X_{%
\mathbf{k}^{^{\prime }}}$'s; in general, the solution is complicated.\
Fortunately, in the thermodynamic limit, it becomes simpler since the
contribution of each single $X_{\mathbf{k}^{^{\prime }}}$ to $\Delta 
\overline{\xi }_{k}$ becomes negligible, so that the sum in (\ref{37}) can
be extended to include $\mathbf{k}^{^{\prime }}=\mathbf{k}$ and become
independent of $\mathbf{k}$.\ This is because, as shown by definition (\ref
{31}), the coupling constant $\alpha $ tends to zero when the volume becomes
infinite, while the correction $\Delta \overline{\xi }_{k}$ remains finite
(in the same limit, the discrete sum becomes an integral which contains a
density of states which is proportional to the volume).\ In fact, for the
contribution of a single $X_{\mathbf{k}^{^{\prime }}}$ to remain
significant, the level in question has to get a population which is
proportional to the volume (extensive).

In what follows we will therefore replace all the $\Delta \overline{\xi }%
_{k} $'s by the same value $\Delta \overline{\xi }$ defined by: 
\begin{equation}
\Delta \overline{\xi }=\alpha \sum_{\mathbf{k}}X_{\mathbf{k}}=\frac{\alpha 
\mathcal{V}}{8\pi ^{3}}\int d^{3}\mathbf{k\,\,}X(\mathbf{k})  \label{44}
\end{equation}
This is possible as long as the system is not Bose condensed; but, if one
level $\mathbf{k}_{0}$ were condensed, one would have to add its
contribution $\alpha X_{\mathbf{k}_{0}}$ to this integral.

\subsection{Graphical solution\label{graphical}}

We can solve the coupled equations by a self-consistent procedure: first we
consider $\Delta \overline{\xi }$ as given, then use its value to calculate
the $X_{\mathbf{k}}$'s, from which we can obtain the sum (we assume that the
system is not condensed): 
\begin{equation}
\frac{\alpha \mathcal{V}}{8\pi ^{3}}\int d^{3}\mathbf{k\,\,}X(\mathbf{k})=4%
\frac{a}{\lambda _{T}}J(\Delta \overline{\xi })  \label{47}
\end{equation}
where $J$ is the dimensionless integral: 
\begin{equation}
J=\left( \frac{\lambda _{T}}{2\pi }\right) ^{3}\int d^{3}\mathbf{k\,\,}X(%
\mathbf{k})  \label{49}
\end{equation}
Finally, we close the system by writing that the integral (\ref{47}) is
nothing but $\Delta \overline{\xi }$, which is equivalent to writing the
relation: 
\begin{equation}
J(\Delta \overline{\xi })=\frac{\lambda _{T}}{4a}\,\Delta \overline{\xi }
\label{49bis}
\end{equation}
and leads to a graphical solution of the self-consistent equations.

Depending on the context, it will be convenient to express $\Delta \overline{%
\xi }$ either in terms of a change $\Delta \overline{\mu }$ of the chemical
potential: 
\begin{equation}
\Delta \overline{\mu }=-\beta ^{-1}\log \left( 1-\Delta \overline{\xi }%
\right)   \label{46}
\end{equation}
or of a change of the fugacity $\Delta z$ defined by: 
\begin{equation}
\Delta \overline{z}=z\left[ \text{e}^{-\beta \Delta \overline{\mu }}-1\right]
=-z\Delta \overline{\xi }  \label{46bis}
\end{equation}
but, needless to say, these three quantities remain essentially the same
parameter.\ For an uncondensed level, (\ref{38}) then provides: 
\begin{equation}
X_{\mathbf{k}}=\frac{1}{\text{e}^{\beta \widetilde{e}_{k}}-1+\Delta 
\overline{\xi }}=\frac{1}{1-\Delta \overline{\xi }}\,\,\frac{1}{\text{e}%
^{\beta \left( \widetilde{e}_{k}+\Delta \overline{\mu }\right) }-1}
\label{45}
\end{equation}
From (\ref{45}) and (\ref{49}) we now get: 
\begin{equation}
J=\frac{1}{1-\Delta \overline{\xi }}\times g_{3/2}\left[ z\left( 1-\Delta 
\overline{\xi }\right) \right]   \label{50}
\end{equation}
where $g_{3/2}(z)$ is defined as usual by: 
\begin{equation}
g_{3/2}(z)=\sum_{l=1}^{\infty }\frac{z^{l}}{l^{3/2}}  \label{g}
\end{equation}
This function is defined only for values of $z$ ranging from $0$ to $1$.
Therefore, when the value of $\mu $ is fixed, the integral $J$ is defined
only for values of $\Delta \overline{\xi }$ ranging from the minimum initial
value $(1-e^{-\beta \mu })$ up to infinity.\ When $\Delta \overline{\xi }$
takes this minimum value, the value of $J$ goes to its maximum: 
\begin{equation}
J_{\max .}=\frac{g_{3/2}(1)}{1-\Delta \overline{\xi }}=\frac{2.61..}{%
1-\Delta \overline{\xi }}  \label{52}
\end{equation}
but, if one added the discrete contribution of the condensed ground state to
the integral, the sum could take any value beyond $J_{\max .}$ for this
particular value of $\Delta \overline{\xi }$, the difference being
proportional to the ground state population. Figure 10 shows plots of $J$ as
a function of $\Delta \overline{\xi }$, the chemical potential $\mu $ being
considered as a parameter; when $\mu $ varies, according to (\ref{52}), the
point of coordinates $(\Delta \overline{\xi },J_{\max .})$ moves on a
hyperbola (shown with a broken line).\FRAME{dtbpFU}{11.6289cm}{7.7233cm}{0pt%
}{\Qcb{Fig.\ 10: Plots of the function $J(\protect\mu ,\Delta \overline{%
\protect\xi })$ as a function of $\Delta \overline{\protect\xi }$ for
various values of the chemical potential $\protect\mu $ \ ($\protect\mu _{1}<%
\protect\mu _{2}<\protect\mu _{3}<\protect\mu _{4}$); for each value of the
chemical potential, the solution of the coupled equations is given by the
intersection point with the straight line $J=\Delta \overline{\protect\xi }%
\protect\lambda _{T}/4a$ (point shown with a circle in the figure). The
dotted line corresponds to the hyperbolic trajectory of the singular point;
the critical point ($\protect\mu =\protect\mu _{4}$) is obtained by
intersecting this hyperbola with the same straight line (point shown with a
square in the figure).}}{\Qlb{fg10}}{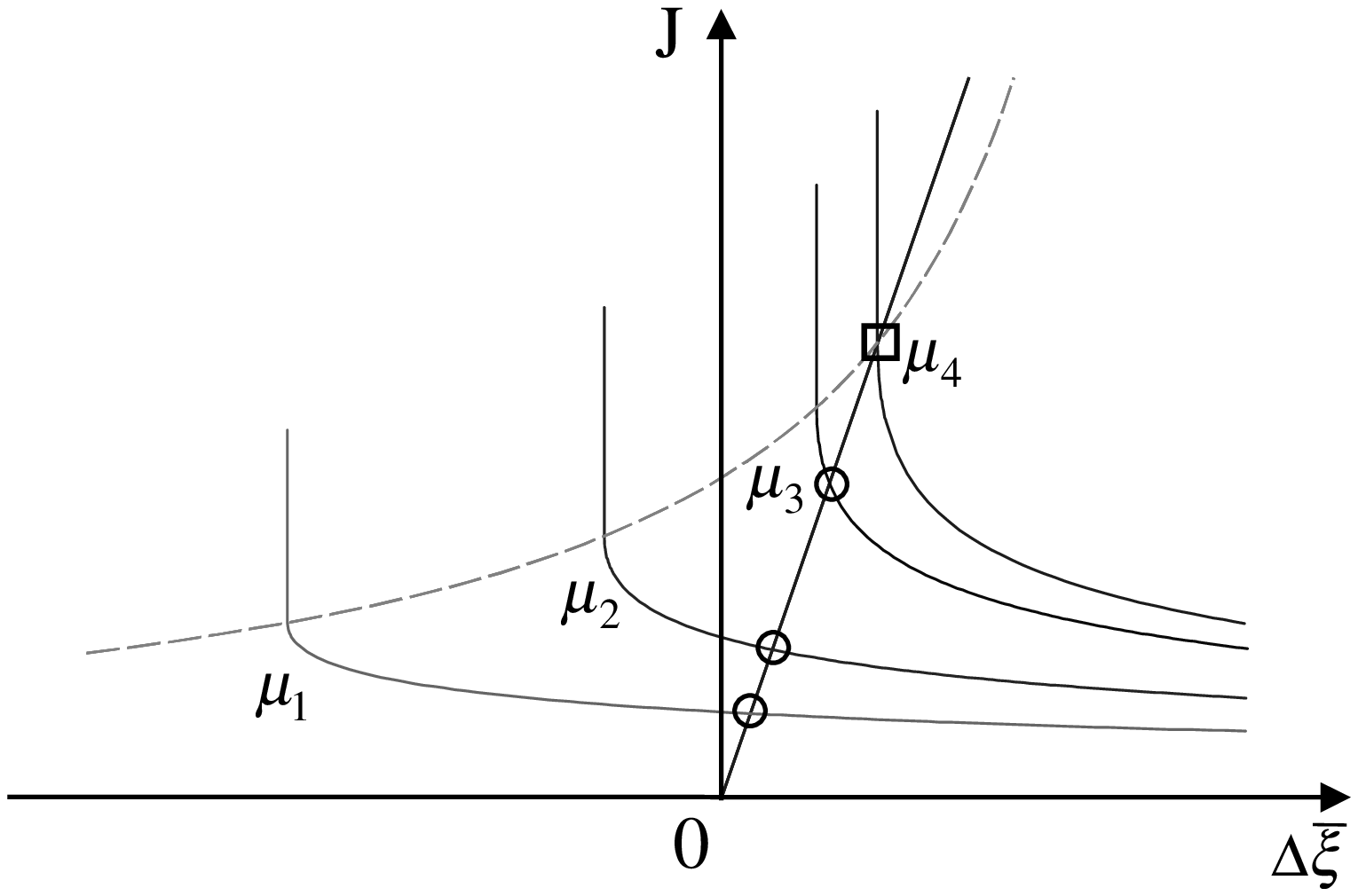}{\special{language "Scientific
Word";type "GRAPHIC";maintain-aspect-ratio TRUE;display "USEDEF";valid_file
"F";width 11.6289cm;height 7.7233cm;depth 0pt;original-width
7.4884in;original-height 4.9649in;cropleft "0";croptop "1";cropright
"1";cropbottom "0";filename 'fg10.eps';file-properties "XNPEU";}}

For any given value of $\mu $, equation (\ref{49bis}) indicates that the
value of $\Delta \overline{\xi }$ is obtained by intersecting the curve
corresponding to $J$ with a straight line going through the origin with
slope $\lambda _{T}/4a$; the intersection point is shown by a circle in the
figure. The construction shows that, as long as $\mu $ is very negative
(classical region), $\Delta \overline{\xi }$ remains only a small
correction; but the situation changes when $\mu $ approaches zero, and more
and more significant corrections are introduced, provided of course the
dilution parameter $a/\lambda _{T}$ is not too small.\ At some critical
positive value $\overline{\mu }_{cr.}$, the value of $\Delta \overline{\xi }$
reaches its limit, and the system undergoes Bose-Einstein condensation%
\footnote{%
The geometrical construction implies that this transition point can be
reached only the value of the dilution parameter $a/\lambda _{T}$ is
sufficiently small; a simple calculation shows that this parameter should be
smaller than $\left[ 4g_{3/2}(1)\right] ^{-1}$ (which is close to 10\%).\ If 
$a/\lambda _{T}$ was larger, the trajectory of the point with coordinates ($%
\Delta \overline{\xi }_{\min .}$, $J_{\max .}$) would pass above the
straight line without ever crossing it; but $a/\lambda _{T}$ is also the
parameter which determines the validity of the low energy approximation that
we have made in the treatment of binary interactions, so that this situation
would simply be beyond the domain of validity of the calculation.}; this
value is merely obtained by intersecting the hyperbola with the straight
line, which provides: 
\begin{equation}
\,\Delta \overline{\xi }_{cr.}=\frac{4a}{\lambda _{T}}\,\frac{2.61..}{%
1-\Delta \overline{\xi }_{cr.}}\simeq 2.61..\frac{4a}{\lambda _{T}}
\label{52bis}
\end{equation}
Since at the transition point $\mu -\Delta \overline{\mu }=0$, this
corresponds to the following value for the chemical potential: 
\begin{equation}
\mu =\overline{\mu }_{cr.}=\Delta \overline{\mu }_{cr.}\simeq \frac{10.44..}{%
\beta }\,\,\frac{a}{\lambda _{T}}  \label{53}
\end{equation}

Finally, equation (\ref{42}) shows that the number of particles as a
function of $\mu $ is given by: 
\begin{equation}
N=\frac{\mathcal{V}}{8\pi ^{3}}\int d^{3}\mathbf{k\,\,}\rho _{\mathbf{k}}=%
\frac{\mathcal{V}}{8\pi ^{3}}\int d^{3}\mathbf{k\,\,}f_{1}(k;\mu -\Delta 
\overline{\mu })  \label{54}
\end{equation}
At the critical value of the chemical potential, $\mu =\Delta \overline{\mu }
$, and the degeneracy parameter $n\lambda _{T}^{3}$ is thus predicted to
have exactly the same value 2.61... as for the ideal gas. Therefore, what we
find within this Ursell-Dyson approximation is a relatively straightforward
result: the only effect of the repulsive interactions is to change the
effective value of the chemical potential, which allows positive values of $%
\mu $ to be reached; otherwise no change (velocity distribution, critical
degeneracy parameter) is introduced.

\section{Beyond mean-field: velocity dependent effects}

We will now go beyond the Ursell-Dyson approximation and study which new
effects are introduced by this more accurate treatment of the problem; we
will see that qualitatively new effects are indeed introduced, which are
strongly velocity-dependent.

\subsection{A new set of equations}

When (\ref{21}) is added to the right hand side of (\ref{18}), a new feature
immediately appears: in (\ref{21}), a momentum transfer $\mathbf{q}$ can now
take place, introducing off-diagonal matrix elements\footnote{%
The matrix elements of $U_{2}$ obey a momentum selection rule, but none for
energy conservation - see for instance Appendix B.} of $U_{2}$. In order to
simplify the calculations, we will nevertheless ignore their dependence on $%
\mathbf{q}$ and merely assume that: 
\begin{equation}
z^{2}<\mathbf{k},\mathbf{k}^{^{\prime }}\mid U_{2}^{S}\mid \mathbf{k}+%
\mathbf{q},\mathbf{k}^{^{\prime }}-\mathbf{q}>=-2\frac{\lambda _{T}^{2}}{%
\mathcal{V}}\,\,\text{e}^{-\frac{\beta }{2}\left[ \widetilde{e}_{\mathbf{k}}+%
\widetilde{e}_{\mathbf{k}^{^{\prime }}}+\widetilde{e}_{\mathbf{k}+\mathbf{q}%
}+\widetilde{e}_{\mathbf{k}^{^{\prime }}-\mathbf{q}}\right] }\times a
\label{55}
\end{equation}
In appendix C we discuss these off-diagonal matrix elements and show that
this is a reasonable approximation; but it is not essential. Equation (\ref
{34}) is then replaced by: 
\begin{equation}
\alpha \sum_{\mathbf{k}^{^{\prime }}}X_{\mathbf{k}^{^{\prime }}}\left[ 1-%
\frac{\alpha }{2}\sum_{\mathbf{q}}X_{\mathbf{k}+\mathbf{q}}X_{\mathbf{k}%
^{^{\prime }}-\mathbf{q}}\right] X_{\mathbf{k}}-\xi _{k}X_{\mathbf{k}}-1=0
\label{56}
\end{equation}
while the generalization to the more complete equation (\ref{23}) would lead
to: 
\begin{equation}
\alpha \sum_{\mathbf{k}^{^{\prime }}}X_{\mathbf{k}^{^{\prime }}}\left[ 1+%
\frac{\alpha }{2}\sum_{\mathbf{q}}X_{\mathbf{k}+\mathbf{q}}X_{\mathbf{k}%
^{^{\prime }}-\mathbf{q}}\right] ^{-1}X_{\mathbf{k}}-\xi _{k}X_{\mathbf{k}%
}-1=0  \label{57}
\end{equation}
We note in passing that, at each ``scale of the ladder'' contained in the
diagrams, momentum conservation takes place, which is why the new sum
extends over one index $q$ only.

In the absence of condensation, the new term becomes an integral with a
factor (density of states) proportional to the volume $\mathcal{V}$, which
cancels the $\mathcal{V}^{-1}$ contained in $\alpha $; the result is then
independent of the volume, exactly as in the Ursell-Dyson approximation.\ On
the other hand, it contains an extra factor $a/\lambda _{T}$ (dilution
parameter), which ensures that the additional correction is smaller than
that considered in the first approximation.\ Now, if the ground state is
condensed, $X_{0}$ becomes proportional to the volume; if $\mathbf{k}+%
\mathbf{k}^{^{\prime }}\neq 0$, the two indices $\mathbf{k}+\mathbf{q}$ and $%
\mathbf{k}^{^{\prime }}-\mathbf{q}$ never vanish for the same value of $%
\mathbf{q}$ so that $X_{0}$ now adds two constants to the integral; if $%
\mathbf{k}+\mathbf{k}^{^{\prime }}=0$, the two indices vanish simultaneously
when $\mathbf{q}=-\mathbf{k}$, so that one gets a contribution proportional
to the volume.\ In the latter case, the sum diverges in the thermodynamic
limit, a situation which never occurred in the Ursell-Dyson approximation;
this is a first indication of a special coupling between opposite values of
the momentum in the presence of condensation.

If only one $X_{\mathbf{k}}$ varies, while all the other are kept constant,
the value of $X_{\mathbf{k}}$ is obtained as the solution of an algebraic
equation; to write it explicitly, it is convenient to expand (\ref{56}) term
by term; when $\mathbf{k}^{^{\prime }}\neq \mathbf{k}$, we obtain: 
\begin{equation}
\alpha X_{\mathbf{k}}\sum_{\mathbf{k}^{^{\prime }}\neq \mathbf{k}}X_{\mathbf{%
k}^{^{\prime }}}\left[ 1-\alpha X_{\mathbf{k}}X_{\mathbf{k}^{^{\prime }}}-%
\frac{\alpha }{2}\sum_{\mathbf{q}\neq 0,\mathbf{k}\neq \mathbf{k}^{^{\prime
}}}X_{\mathbf{k}+\mathbf{q}}X_{\mathbf{k}^{^{\prime }}-\mathbf{q}}\right]
\label{58}
\end{equation}
while, if $\mathbf{k}^{^{\prime }}=\mathbf{k}$, the result is changed into: 
\begin{equation}
\alpha \left( X_{\mathbf{k}}\right) ^{2}\left[ 1-\frac{\alpha }{2}\left( X_{%
\mathbf{k}}\right) ^{2}-\frac{\alpha }{2}\sum_{\mathbf{q}\neq 0}X_{\mathbf{k}%
+\mathbf{q}}X_{\mathbf{k}-\mathbf{q}}\right]  \label{59}
\end{equation}
Adding these contributions provides the equation: 
\begin{equation}
-\frac{\alpha ^{2}}{2}\left( X_{\mathbf{k}}\right) ^{4}+\alpha \left( X_{%
\mathbf{k}}\right) ^{2}\left[ 1-\frac{\alpha }{2}\sum_{\mathbf{q}\neq 0}X_{%
\mathbf{k}+\mathbf{q}}X_{\mathbf{k}-\mathbf{q}}-\alpha \sum_{\mathbf{k}%
^{^{\prime }}}\left( X_{\mathbf{k}^{^{\prime }}}\right) ^{2}\right] -\xi _{%
\mathbf{k}}^{^{\prime }}X_{\mathbf{k}}-1=0  \label{60}
\end{equation}
where $\xi _{\mathbf{k}}^{^{\prime }}$ is now defined by: 
\begin{equation}
\xi _{\mathbf{k}}^{^{\prime }}=\xi _{\mathbf{k}}-\alpha \sum_{\mathbf{k}%
^{^{\prime }}}X_{\mathbf{k}^{^{\prime }}}+\frac{\alpha ^{2}}{2}\sum_{\mathbf{%
k}^{^{\prime }}\neq \mathbf{k}}X_{\mathbf{k}^{^{\prime }}}\sum_{\mathbf{q}%
\neq 0,\mathbf{k\neq k}^{^{\prime }}}X_{\mathbf{k}+\mathbf{q}}X_{\mathbf{k}%
^{^{\prime }}-\mathbf{q}}  \label{61}
\end{equation}
These results generalize (\ref{34bis}), (\ref{36}) and (\ref{37}); they are
no longer second degree equations, since they contain higher degree terms.\
Nevertheless, all these terms contain one extra power of $\alpha $, which is
inversely proportional to the volume, so that in the thermodynamic limit
they remain negligible as long as $X_{\mathbf{k}}$ is not proportional (at
least) to the square root of the volume - actually, exactly as was already
the case for the quadratic term in (\ref{34bis}). Under these conditions,
the situation remains essentially similar to that discussed in the preceding
section, and $X_{\mathbf{k}}$ is determined by the linear terms only: 
\begin{equation}
X_{\mathbf{k}}=-\frac{1}{\xi _{\mathbf{k}}^{^{\prime }}}  \label{62}
\end{equation}
The only difference is that $\xi _{\mathbf{k}}^{^{\prime }}$ is now given by
(\ref{61}), which contains a correction proportional to $\alpha ^{2}$.

If we change sums into integrals, we get: 
\begin{equation}
\xi _{\mathbf{k}}^{^{\prime }}=\xi _{\mathbf{k}}-\Delta \overline{\xi }%
+\delta \xi _{k}  \label{63}
\end{equation}
where $\Delta \overline{\xi }$ is defined as above by: 
\begin{equation}
\Delta \overline{\xi }=4\frac{a}{\lambda _{T}}\left( \frac{\lambda _{T}}{%
2\pi }\right) ^{3}\int d^{3}\mathbf{k}^{^{\prime }}\,\,X(\mathbf{k}%
^{^{\prime }})  \label{64}
\end{equation}
while the additional (second order) correction $\delta \xi _{k}$ is given
by: 
\begin{equation}
\delta \xi _{k}=8\left( \frac{a}{\lambda _{T}}\right) ^{2}\left( \frac{%
\lambda _{T}}{2\pi }\right) ^{6}\int d^{3}\mathbf{k}^{^{\prime }}\mathbf{\,}%
\,X(\mathbf{k}^{^{\prime }})\int d^{3}\mathbf{q\,\,}X(\mathbf{k}+\mathbf{q)}%
X(\mathbf{k}^{^{\prime }}-\mathbf{q)}  \label{65}
\end{equation}
A new feature which appears in this correction is a dependence on $\mathbf{k}
$, which will introduce velocity-dependent effects; this can be seen better
by rewriting the second integral in the form: 
\begin{equation}
\int d^{3}\mathbf{q}^{^{\prime }}\mathbf{\,\,\,}X(\frac{\mathbf{k+k}%
^{^{\prime }}}{2}+\mathbf{q}^{^{\prime }})\,\,X(\frac{\mathbf{k+k}^{^{\prime
}}}{2}-\mathbf{q}^{^{\prime }})  \label{66}
\end{equation}
which appears as a convolution integral of the momentum distribution with
itself, with a maximum when $\mathbf{k+k}^{^{\prime }}=0$.\ As a
consequence, while $X_{\mathbf{k}}$ is coupled through $\delta \xi _{k}$ to
all other $X_{\mathbf{k}^{^{\prime }}}$'s, the maximum coupling takes place
with $X_{-\mathbf{k}}$, which then implies that $\delta \xi _{k}$ is a
larger correction near the center of the velocity distribution (because the
preferred coupling occurs with $X_{\mathbf{k}^{^{\prime }}}$'s which are
large) than in its wings (where velocity classes are preferentially coupled
to smaller $X_{\mathbf{k}^{^{\prime }}}$'s).

\subsection{A shift in the transition parameter\label{shift}}

As in \S\ \ref{mean}, we have to solve equations consistently since the $X_{%
\mathbf{k}}$'s depend on the corrections $\Delta \overline{\xi }$ and $%
\delta \xi _{k}$ which, in turn, depend on the $X_{\mathbf{k}}$'s, a
complicated problem in general.\ Nevertheless, we can already get a general
idea of the changes introduced by the new corrections by assuming for a
moment that they are known, or at least just the correction $\delta \xi _{0}$
for the ground state.\ Figure 11 then summarizes the comparison between the
Ursell-Dyson approximation and the theory including $\delta \xi _{0}$. The
full line shows the variations of $\Delta \overline{\mu }$ as a function of $%
\mu $, which at $\mu =\overline{\mu }_{cr.}$ reaches tangentially the
straight line corresponding to the condensation condition $\mu +\Delta 
\overline{\mu }=0$; we note that, when $\mu \rightarrow \overline{\mu }%
_{cr.} $ (by lower values), $\Delta \overline{\mu }$ remains a function of
the chemical potential which keeps a continuous derivative\footnote{%
Figure 10 shows that the slope of the curve giving the variations of $J$ as
a function of $\Delta \overline{\xi }$ becomes infinite when the curve
connects to the vertical line - no discontinuity takes place in the
derivative of $\Delta \overline{\xi }$ as a function of $\mu $.}. The broken
line illustrates the changes to this construction introduced by $\delta \xi
_{0}$, or equivalently by the correction $\delta \mu _{0}$ defined by
analogy with (\ref{46}) (and for any value of $\mathbf{k}$) as: 
\begin{equation}
\delta \mu _{k}-\Delta \overline{\mu }=\beta ^{-1}\log \left( 1-\Delta 
\overline{\xi }+\delta \xi _{k}\right)  \label{67}
\end{equation}
The new transition point occurs when the total effective chemical potential
of the ground state vanishes: 
\begin{equation}
\mu -\Delta \overline{\mu }+\delta \mu _{0}=0  \label{67bis}
\end{equation}
which corresponds to a lower value $\mu _{cr.}$ of the chemical potential,
and no longer to a tangential contact. An interesting feature is that the
square root behavior of $\Delta \overline{\mu }-\mu $ near $\mu =\overline{%
\mu }_{cr.}$ may introduce a non-analytical behavior of the solution; in
appendix D we write down in more detail the local expansion of $\Delta 
\overline{\mu }$ and show, for instance, why the correction to $\mu _{cr.}$
is second order in $a$, despite this square root behavior. Qualitatively, we
can already guess that the critical value of the degeneracy parameter will
be lowered in the process; this is because the excited levels get a
correction $\delta \xi _{k}$ which is smaller than $\delta \xi _{0}$, and
therefore experience at the new transition point an effective chemical
potential which is more negative than it was in the Ursell-Dyson approach (a
differential effect); but we now evaluate this correction more qualitatively.%
\FRAME{dtbpFU}{4.3076in}{2.4846in}{0pt}{\Qcb{Fig.\ 11: Variations as a
function of $\protect\mu $ of the mean-field correction $\Delta \overline{%
\protect\mu }$ to the chemical potential (full line); the corresponding
critical value is $\overline{\protect\mu }_{cr.}$. A mored detailed
treatment includes the extra correction $-\protect\delta \protect\mu _{0}$
(broken line) and reduces the critical value to a new value $\protect\mu 
_{cr.}$. }}{\Qlb{fg11}}{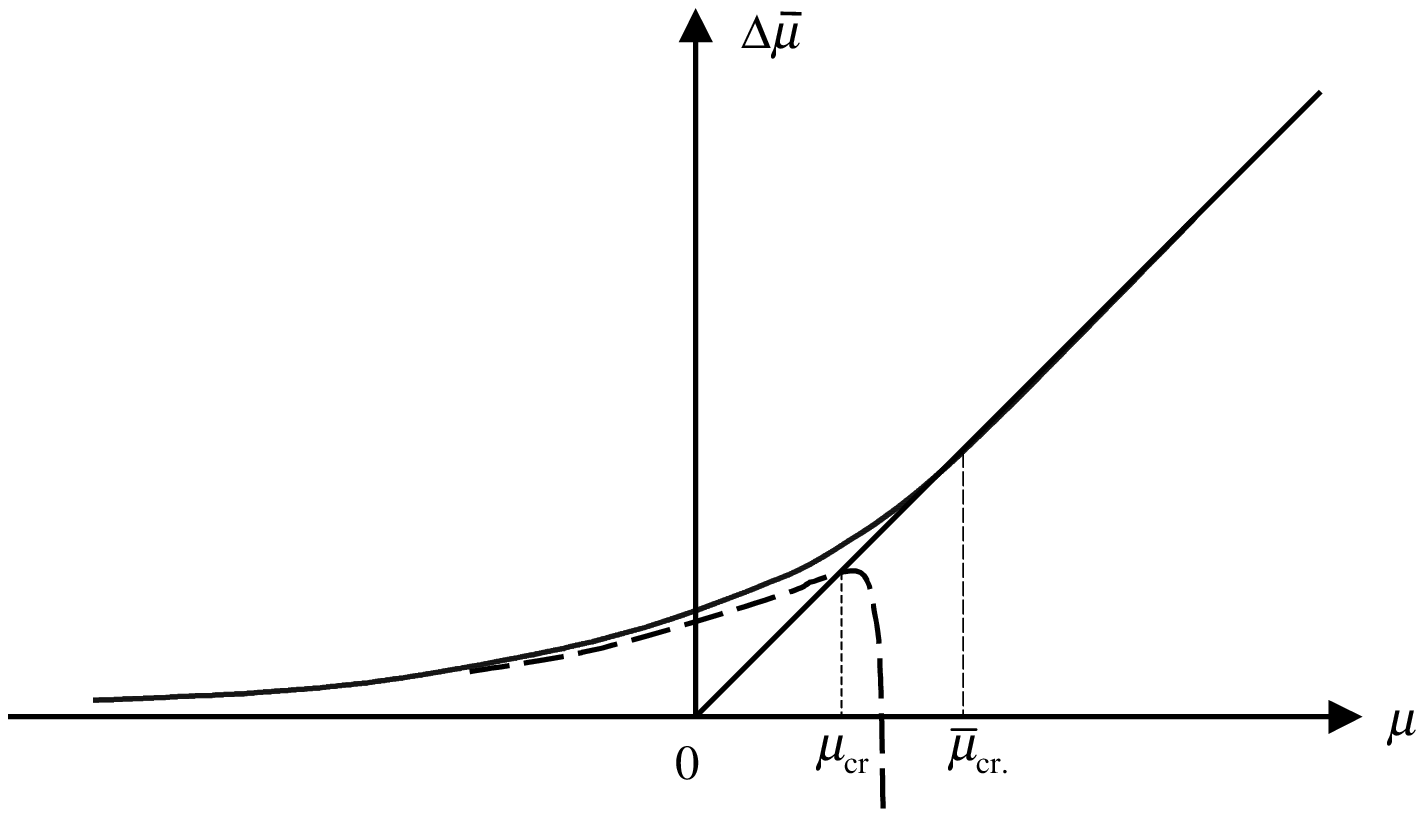}{\special{language "Scientific Word";type
"GRAPHIC";maintain-aspect-ratio TRUE;display "USEDEF";valid_file "F";width
4.3076in;height 2.4846in;depth 0pt;original-width 6.794in;original-height
3.9072in;cropleft "0";croptop "1";cropright "1";cropbottom "0";filename
'fg11.eps';file-properties "XNPEU";}}

A graphical solution of the consistent equations would be more complicated
than in the preceding section, mostly because one single parameter $\Delta 
\overline{\xi }$ is no longer sufficient to characterize the effect of the
interactions on all velocity classes. We can nevertheless resort to a
``first $\delta \xi $ iteration'', namely calculate the $\delta \xi _{k}$'s
from (\ref{65}) by assuming that the $X_{\mathbf{k}}$ still have the values
that they had in the Ursell-Dyson approximation, and then use this
correction to calculate new $X_{\mathbf{k}}$'s as well as new populations.\
Of course, in theory, this should be only the initial step of an iteration
procedure, which should be repeated until convergence is obtained, but for
the moment we limit ourselves to this first approximation.

Under these conditions, (\ref{65}) becomes: 
\begin{equation}
\delta \xi _{k}\simeq 4\left( \frac{a}{\lambda _{T}}\right) \left( \frac{%
\lambda _{T}}{2\pi }\right) ^{3}\text{e}^{\beta \Delta \overline{\mu }}\int
d^{3}\mathbf{k}^{^{\prime }}\mathbf{\,}\,\,\,f_{1}(\mathbf{k}^{^{\prime
}};\mu -\Delta \overline{\mu })\,\,J(\mathbf{k}+\mathbf{k}^{^{\prime }};\mu
-\Delta \overline{\mu })  \label{67-1}
\end{equation}
where the integral $J$ is defined by: 
\begin{equation}
J(\mathbf{K};\mu -\Delta \overline{\mu })=2\left( \frac{a}{\lambda _{T}}%
\right) \left( \frac{\lambda _{T}}{2\pi }\right) ^{3}\text{e}^{2\beta \Delta 
\overline{\mu }}\int d^{3}\mathbf{q\,\,\,}f_{1}\mathbf{(}\frac{\mathbf{K}}{2}%
\mathbf{\mathbf{+q};}\mu -\Delta \overline{\mu }\mathbf{)\,\,\,}f_{1}\mathbf{%
(}\frac{\mathbf{K}}{2}\mathbf{-q;}\mu -\Delta \overline{\mu }\mathbf{)\,}
\label{67-2}
\end{equation}
These two integrals can be calculated by expanding $f_{1}$ in series as in
equation (\ref{14bis}); integrating the Gaussian functions provides: 
\begin{equation}
\begin{array}{ll}
\displaystyle J(\mathbf{K};\mu -\Delta \overline{\mu }) & \displaystyle%
=2\left( \frac{a}{\lambda _{T}}\right) \text{e}^{2\beta \Delta \overline{\mu 
}}\sum_{l,m}\left( l+n\right) ^{-3/2}\times \\ 
& \multicolumn{1}{r}{\displaystyle\times \text{exp}\left\{ -\frac{l\,\,n}{%
l\,+\,n}\frac{\lambda _{T}^{2}}{4\pi }K^{2}+\beta \left( l+n\right) \left(
\mu -\Delta \overline{\mu }\right) \right\}}
\end{array}
\label{68}
\end{equation}
and: 
\begin{equation}
\begin{array}{ll}
\displaystyle\delta \xi _{k} & \displaystyle\simeq 8\left( \frac{a}{\lambda
_{T}}\right) ^{2}\text{e}^{3\beta \Delta \overline{\mu }}\sum_{l,n,p}\left[
(l+n)p+l\,\,n\right] ^{-3/2}\times \\ 
& \multicolumn{1}{r}{\displaystyle\times \text{exp}\left\{ -\frac{l\,\,n\,\,p%
}{\left( l+n\right) p+\,\,l\,\,n}\frac{\lambda _{T}^{2}}{4\pi }k^{2}+\beta
\left( l+n+p\right) \left( \mu -\Delta \overline{\mu }\right) \right\}}
\end{array}
\label{69}
\end{equation}

The procedure is therefore the following: we start from an arbitrary value
of $\mu $ and use the above expressions (as well as those of \S \ref
{graphical} for the mean-field values) to calculate the effective chemical
potential (\ref{67}) for each velocity class; as long as this function
remains negative for all values of $\mathbf{k}$, the population $\rho _{%
\mathbf{k}}$ of each level is given by the direct generalization of (\ref{42}%
): 
\begin{equation}
\rho _{\mathbf{k}}=f_{1}(k;\mu -\Delta \overline{\mu }+\delta \mu _{k})
\label{70bis}
\end{equation}
Now, when $\mu $ increases, at some point condition (\ref{67bis}) is met -
see figure 11 - and condensation takes place.\ Figure 12 shows a plot of the
variations of this critical value of $\mu $ as a function of the
dimensionless parameter $a/\lambda _{T}$, with, as a point of comparison,
the dotted straight line obtained from the mean-field value (\ref{53}); we
see that the mean-field theory gives a good approximation of the value of
the chemical potential.\FRAME{dtbpFU}{4.2817in}{3.1964in}{0pt}{\Qcb{Fig.\
12: variations of the critical value of the chemical potential as a function
of the dimensionless interaction parameter $a/\protect\lambda _{T}$ of the
gas. The dotted upper straight line corresponds to the prediction of the
Ursell-Dyson (mean-field) approach, equation (\ref{53}); the lower line
includes the velocity dependent effects, threated within the first iteration
corresponding to equations (\ref{69}).}}{\Qlb{fg12}}{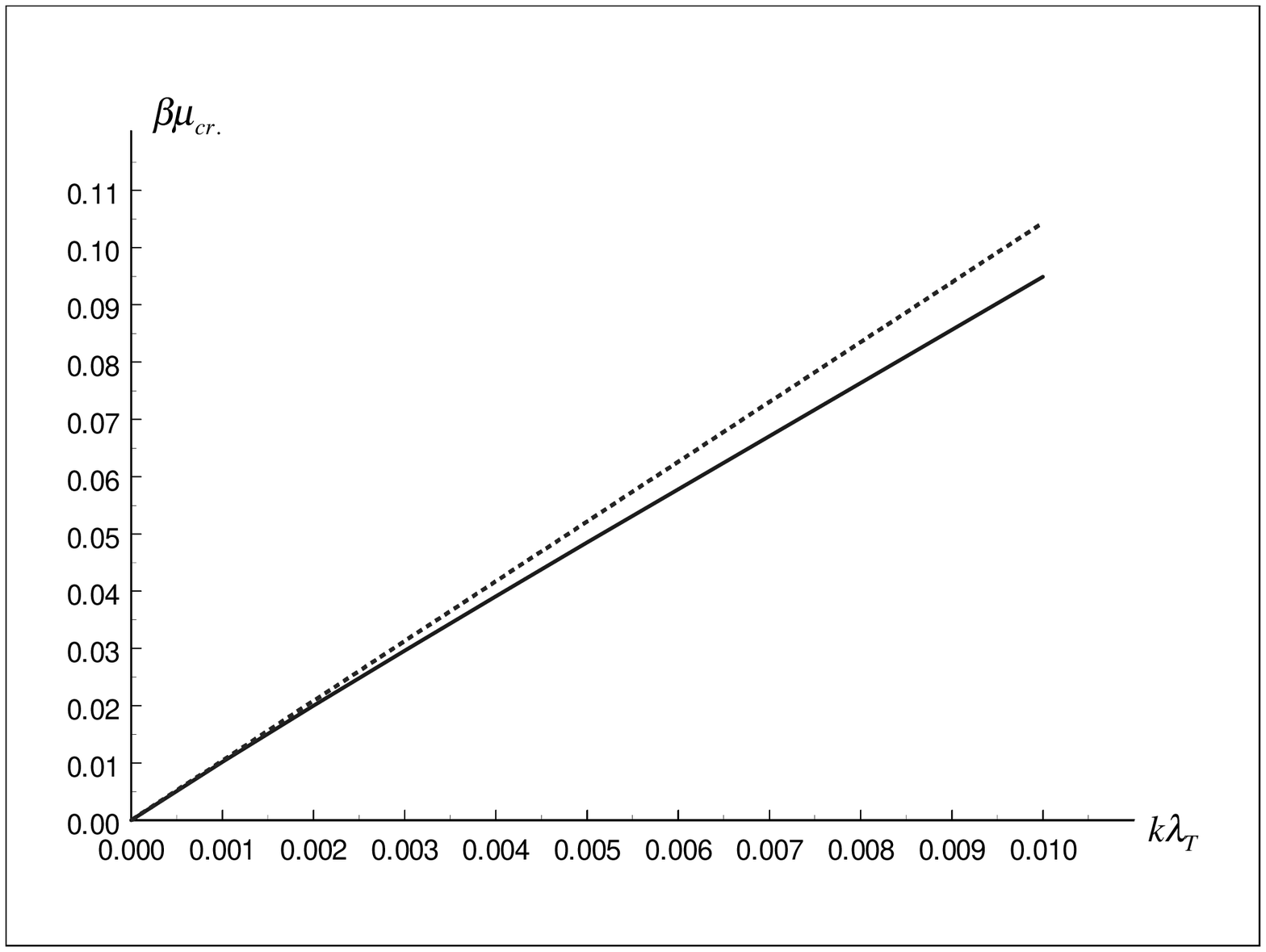}{\special%
{language "Scientific Word";type "GRAPHIC";maintain-aspect-ratio
TRUE;display "USEDEF";valid_file "F";width 4.2817in;height 3.1964in;depth
0pt;original-width 6.9047in;original-height 5.1474in;cropleft "0";croptop
"1";cropright "1";cropbottom "0";filename 'fg12.eps';file-properties
"XNPEU";}}\ For this critical value of $\mu $, and for three values of the
parameter $a/\lambda _{T}$, figure 13 then shows the variations of $\delta
\xi _{k}\simeq \beta \delta \mu _{k}$ as a function of $k$, in units $%
\lambda _{T}^{-1}$ where the width of he thermal velocity distribution $%
k^{2}f_{1}(k,0)$ is about $4$. An obvious feature in the figure is that the
correction $\delta \mu _{k}$ is neither constant (which would lead to an
unchanged critical degeneracy parameter) nor a simple quadratic function of $%
k$ (which would lead to an effective mass effect).\ In fact it has a
structure within the thermal velocity distribution, which resembles a
Gaussian function, and which is narrower, but not by a large factor ($2$ or $%
3$ in this case); this creates a intermediate physical situation so that
quantitative predictions require a somewhat more detailed study of the
perturbations within the velocity profile.\FRAME{dtbpFU}{4.587in}{2.879in}{%
0pt}{\Qcb{Fig.\ 13: variations of the correction to the energy of the atoms
as a function of their momentum, in dimensionless units $k\protect\lambda
_{T}$ (the half-height width of the thermal distribution is then about $4$);
the results are obtained by a first iteration of the non-linear equations
and the value of the chemical potential is adjusted to its critical value
within this approximation.\ The lowest curve corresponds to $a/\protect%
\lambda _{T}=0.002$ and its with is $1.55$; the middle curve to to $a/%
\protect\lambda _{T}=0.003$, and its width $1.75$; the upper curve is for $a/%
\protect\lambda _{T}=0.004$ and its width is $1.95$. This shows that the
velocity-dependent corrections are localized at the center of the velocity
profile.}}{\Qlb{fg13}}{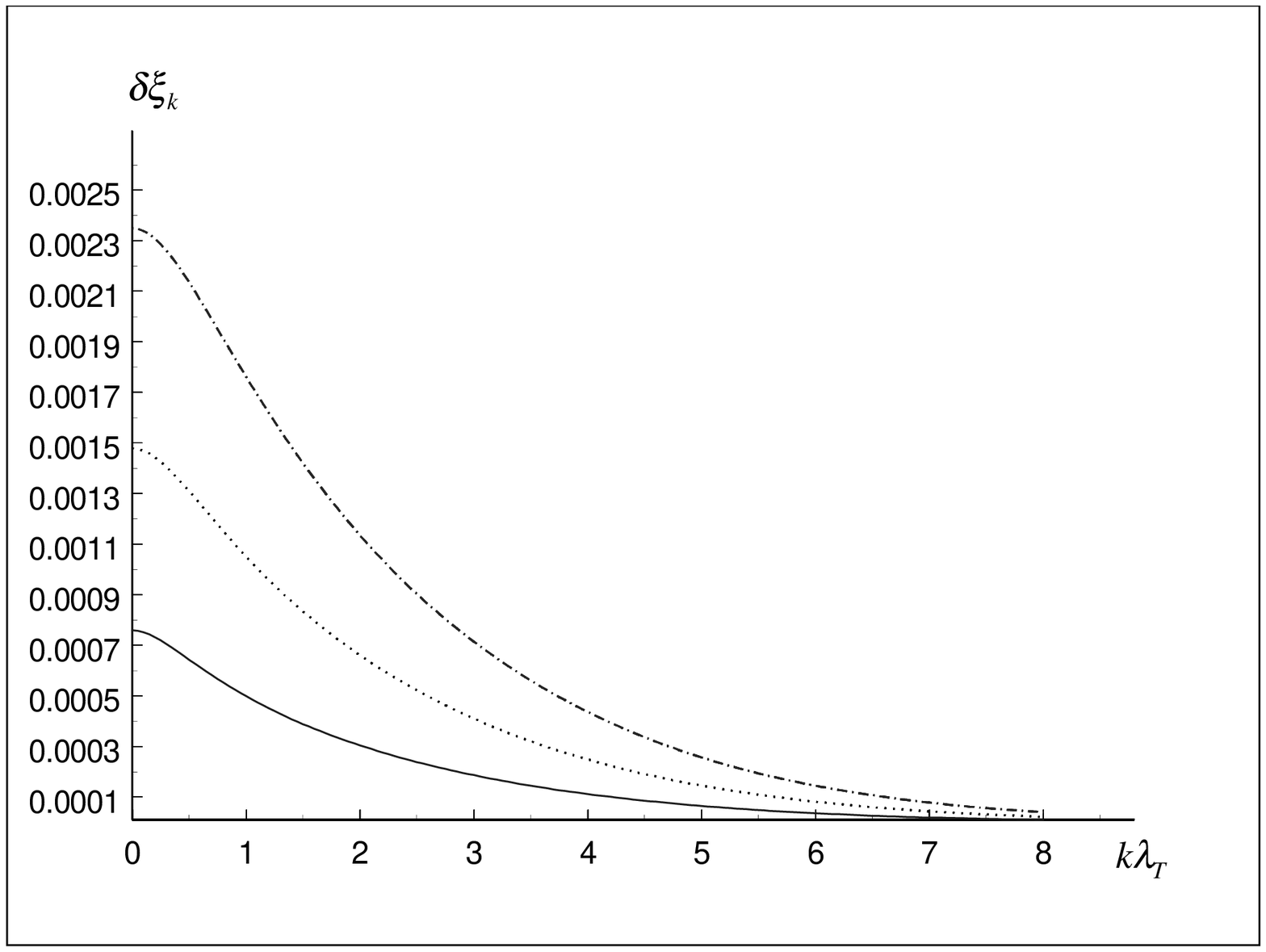}{\special{language "Scientific Word";type
"GRAPHIC";display "USEDEF";valid_file "F";width 4.587in;height 2.879in;depth
0pt;original-width 7.0733in;original-height 3.8674in;cropleft "0";croptop
"1";cropright "1";cropbottom "0";filename 'fg13.eps';file-properties
"XNPEU";}} According to (\ref{67bis}), the expression of the critical
density can be obtained by the following summation over velocities: 
\begin{equation}
n_{cr.}=\frac{1}{(2\pi )^{3}}\int d^{3}\mathbf{k}\,\,f_{1}(\mathbf{k},\mu
=\delta \mu _{0}-\delta \mu _{k})  \label{70}
\end{equation}
This expression was used to calculate numerically the values of the critical
degeneracy parameter $n\lambda _{T}^{3}$ as a function of the dilution
parameter $a/\lambda _{T}$ (also equal to $1.38..\times n^{1/3}a$ near the
transition temperature); the results are shown in figure 14, expressed in
terms of the difference between this critical value and the value 2.612..
for the ideal gas. We see that the integration over velocities gives rise to
a correction which turns out to be, at least approximately, linear in $%
a/\lambda _{T}$.\ As a point of comparison, we also show the results of a
recent path integral quantum Monte-Carlo calculation \cite{Gruter}; the
agreement is satisfactory but not perfect; below we discuss possible
improvements.\FRAME{dtbpFU}{5.0695in}{3.2724in}{0pt}{\Qcb{Fig.\ 14: relative
changes of the degeneracy parameter $n\protect\lambda _{T}^{3}$ as a
function of the density of the gas; the left vertical axis gives the
opposite of this (negative) change, while the right vertical axis
corresponds to the equivalent (positive) change of the critical temperature
at constant density.\ The lower line shows the results of the quantum Monte
Carlo calculations of ref. \protect\cite{Gruter}; \ the upper curve
corresponds to the results of our calculation within the first iteration of
\S\ \ref{shift}; the middle curve to the more calculation discussed at the
end of \S\ \ref{discussion}, where a denominator $1+J(\mathbf{k}+\mathbf{k}%
^{^{\prime }})$ is added within the integral giving $\protect\delta \protect%
\xi _{k}$ (but still with only one iteration of the non linear equations).}}{%
\Qlb{fg14}}{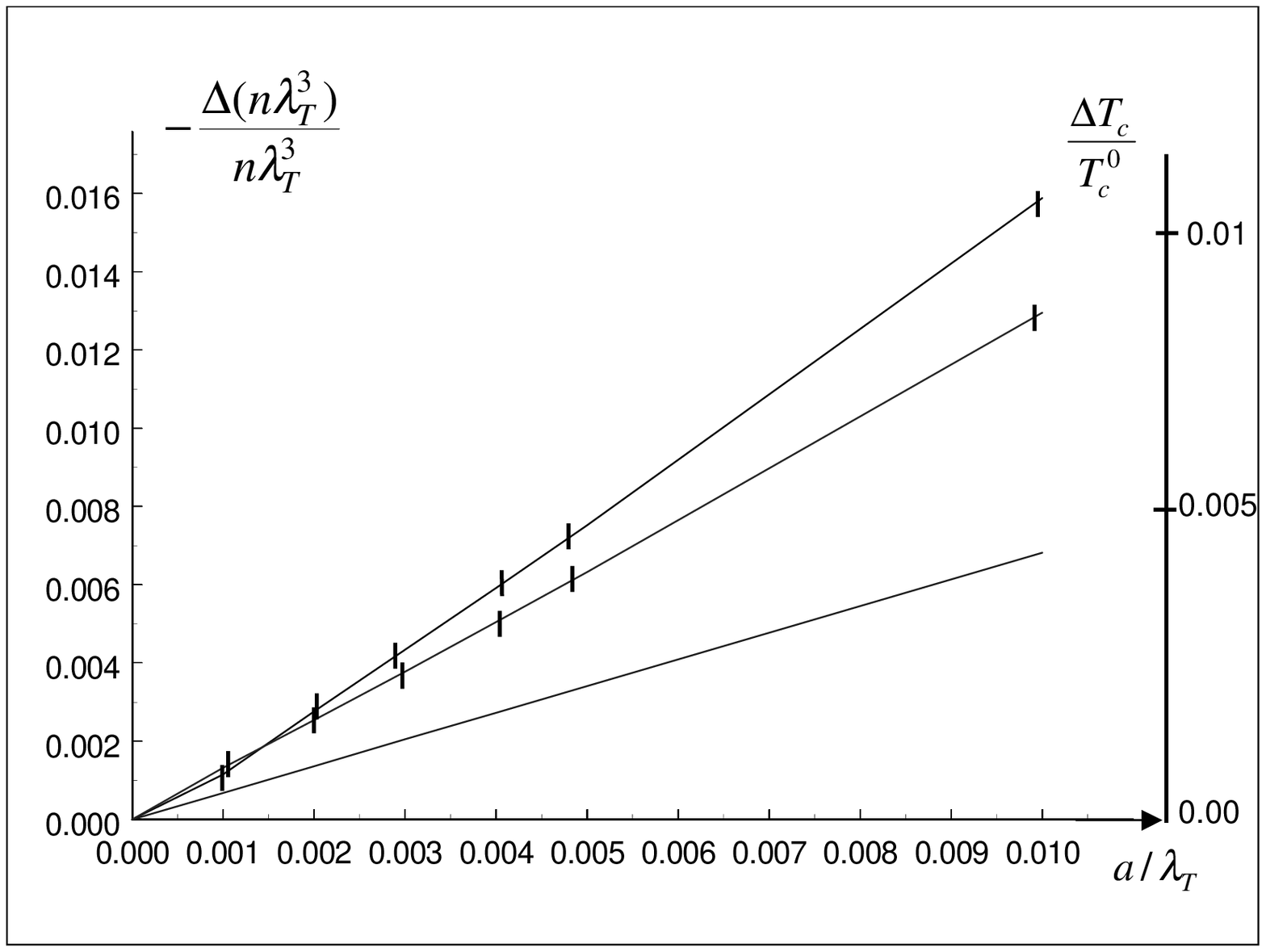}{\special{language "Scientific Word";type
"GRAPHIC";maintain-aspect-ratio TRUE;display "USEDEF";valid_file "F";width
5.0695in;height 3.2724in;depth 0pt;original-width 7.1278in;original-height
4.5913in;cropleft "0";croptop "1";cropright "1";cropbottom "0";filename
'fg14.eps';file-properties "XNPEU";}}

At this point, it becomes possible to discuss in more detail the physical
origin of these results.\ Mathematically, the key role in the mechanism is
played by $\delta \xi _{\mathbf{k}}$, a correction which tends to increase
effective chemical potential of a velocity class $\mathbf{k}$, while $\Delta
\xi $ had just the opposite effect. In other words, the mean-field
approximation tends to overestimate the repulsion between the velocity
classes, so that a correction with the opposite sign is introduced by $%
\delta \xi _{\mathbf{k}}$; the correction is more important for low velocity
classes than for rapid particles.\ Physically, it is indeed natural to
expect that slow particles should be more sensitive to interactions than
fast particles: they can more easily rearrange themselves spatially in order
to minimize their repulsive interactions.\ By contrast, fast particles have
too much kinetic energy to have the same sensitivity to small perturbations;
the rearrangement does not take place, so that they experience the full mean
repulsion $\Delta \overline{\xi }$ (as for uncorrelated particles).\ The
velocity dependence of the correction shown in figure 13; at the critical
temperature, the populations of the low $\mathbf{k}$ levels are almost those
of the corresponding ideal gas, with only a mass renormalization effect
introduced by the curvature of the curve at the origin; but those of higher
velocities levels get almost no correction $\delta \xi _{\mathbf{k}}$, so
that they remain smaller than those of the ideal gas with chemical potential 
$\mu -\Delta \overline{\mu }$.\ Integrating over all velocities provides a
reduced value of the degeneracy parameter $n\lambda _{T}^{3}$.

Our results go in the same direction than the views expressed by
Nozi\`{e}res \cite{Nozieres} concerning the effects of repulsive
interactions in a Bose gas, which were found to stabilize the gas against
smeared condensation (fractionned condensate): figure 12 predicts that the
populations of the low excited levels are indeed lower than they would be in
an ideal gas, which is an indication or a stronger tendency to populate the
ground state exclusively. The physical origin of the effect is nevertheless
rather different: ref. \cite{Nozieres} is based on the use of a mean-field
calculation, where no re-arrangement (or change of the velocity profile) can
take place; here we find that the changes originate from a differential
effect which is essentially beyond mean-field theory. But \cite{Nozieres}
treats a system which is already condensed, while we approach the transition
from the non-condensed phase, so that the physics may well be different in
both cases. We also note that the effective mass effect at the center of
figure 11 is of different nature than that calculated in \cite
{Fetter-Walecka}; in particular, it does not require the potential to have a
finite range to occur (also, it has the opposite sign: we find a smaller
effective mass than the bare mass).

\subsection{ Discussion; more precise calculations\label{discussion}}

The calculations that we have developed are not exact, for two reasons:

(i) we have not solved the coupled equations between the $\delta \xi _{%
\mathbf{k}}$'s and the $X_{\mathbf{k}}$, but only a ``first $\delta \xi $
iteration'' of these equations.

(ii) the initial equation (\ref{56}) is not exact; (\ref{57}) is a better
approximation, but not exact either, and can be completed with a larger
class of diagrams.

When approximation (i) is not made, equations (\ref{67-1}) to (\ref{70}) are
replaced by a system containing, first, the equations providing the $X_{%
\mathbf{k}}$'s: 
\begin{equation}
X_{\mathbf{k}}=\frac{f_{1}\left[ \mathbf{k},z(1-\Delta \overline{\xi }%
+\delta \xi _{\mathbf{k}})\right] }{1-\Delta \overline{\xi }+\delta \xi _{%
\mathbf{k}}}  \label{71}
\end{equation}
and, second, the equations providing the $\delta \xi _{\mathbf{k}}$'s (the
mean repulsion $\Delta \overline{\xi }$ keeps the same expression as in the
mean-field theory): 
\begin{equation}
\delta \xi _{\mathbf{k}}=4\left( \frac{a}{\lambda _{T}}\right) \left( \frac{%
\lambda _{T}}{2\pi }\right) ^{3}\int d^{3}\mathbf{k}^{^{\prime }}\,X_{%
\mathbf{k}^{^{\prime }}}\times J(\mathbf{k}+\mathbf{k}^{^{\prime }})
\label{72}
\end{equation}
with: 
\begin{equation}
J(\mathbf{K})=2\left( \frac{a}{\lambda _{T}}\right) \left( \frac{\lambda _{T}%
}{2\pi }\right) ^{3}\int d^{3}\mathbf{q\,}X_{\mathbf{q+K}/2}\times X_{-%
\mathbf{q}+\mathbf{K}/2}  \label{73}
\end{equation}
We do not intend to solve these coupled equations in detail here, but just
to discuss in general terms what kind of new physical effects they
introduce, as compared to those of the preceding section.\ A first
indication can be obtained by studying what would have been obtained in a
``second $\delta \xi $ iteration''.\ Instead of the expressions (\ref{67-1})
and (\ref{67-2}) containing ordinary Bose-Einstein distributions, we would
have inside the integrals perturbed distributions, with already a relatively
narrow enhancement near the center due to the first iteration.\ The
narrowing effect of the auto-convolution integral, which occurred in the
step, would take place again, so that we would get a still more localized
perturbation at the center of the velocity profile.\ Now, when more and more
iterations are added, narrower and narrower perturbations are expected to be
introduced at the center, so that at the end there is no reason why the
correction to \ the energy should remain quadratic at the origin.\ The
situation is shown in figure 15.\FRAME{dtbpFU}{4.216in}{3.0744in}{0pt}{\Qcb{%
Fig. 15: when more and more iterations are performed in the calculation of
the corrections to the energy of the particles, a sharper and sharper
localized perturbation occurs at low velocities, so that eventually the
spectrum no longer remains quadratic at the origin, but acquires a
dependence in $k^{-3/2}$.}}{}{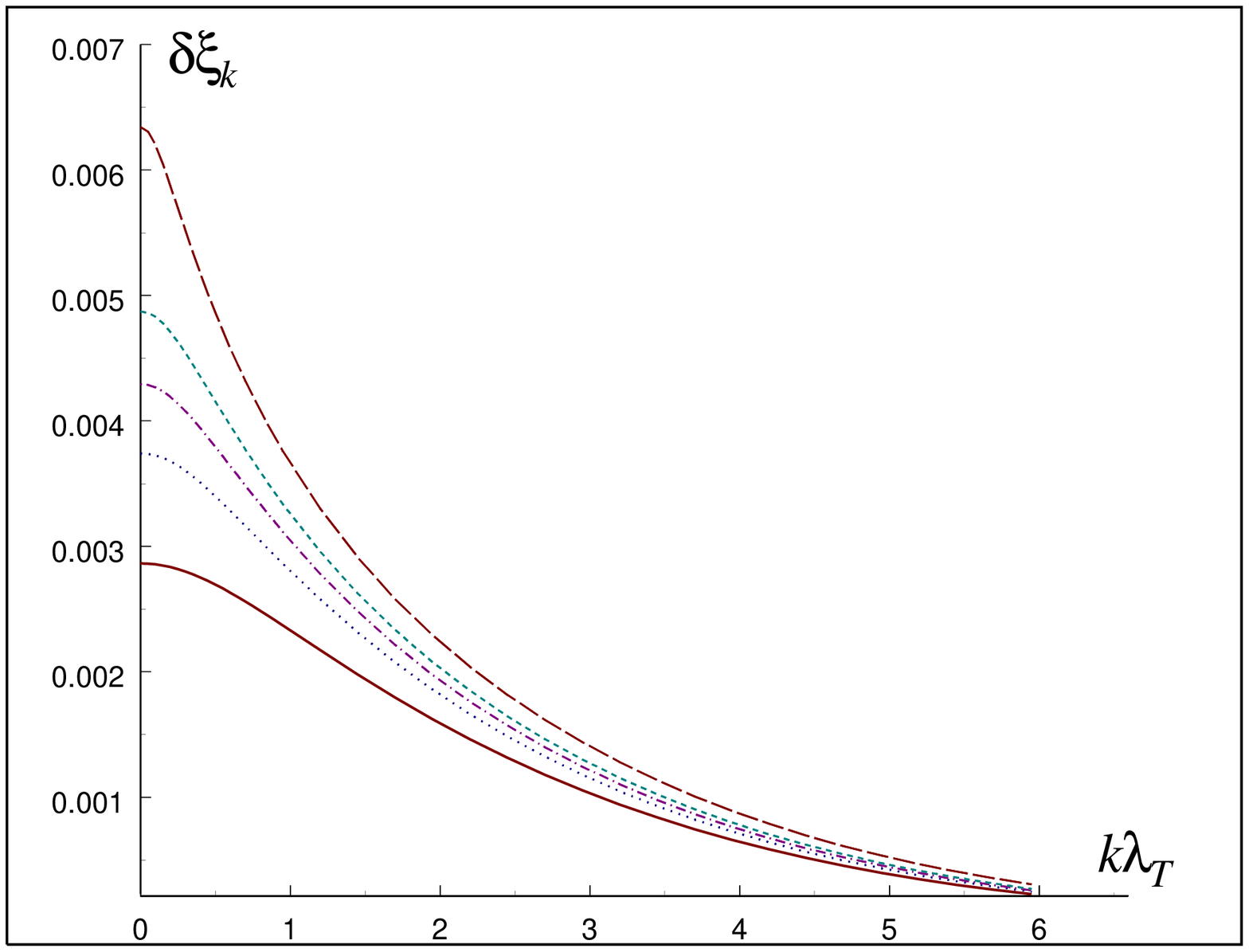}{\special{language "Scientific
Word";type "GRAPHIC";display "USEDEF";valid_file "F";width 4.216in;height
3.0744in;depth 0pt;original-width 7.3362in;original-height 3.7844in;cropleft
"0";croptop "1";cropright "1";cropbottom "0";filename
'fg15.eps';file-properties "NPEU";}}

Actually, this can also be seen from the fact that (\ref{72}) gives a
divergent integral\footnote{%
The first iteration (\ref{67-1}) of (\ref{72}) already diverges, as the
logarithm of $\left| \mu -\Delta \overline{\mu }\right| $ \cite{René}; this
divergence is taken into account in figure 10.} at the critical value of the
chemical potential if the $X_{\mathbf{k}}$'s diverge at the origin as $%
k^{-2} $; such a divergence would be incompatible with the relation $\mu
-\Delta \overline{\mu }+\delta \xi _{0}=0$ at the transition point, as
visible for instance in figure 11.\ In other words, the integral giving $%
\delta \xi _{0}$ plays the role of a restoring force in the problem, and the
solution has to adapt to keep it finite. Assume then that, at the
condensation point, the variations $\delta \xi _{k}$ for small values of $k$
correspond to a spectrum with a power $\alpha $ of the momentum$:$%
\begin{equation}
\delta \xi _{0}-\delta \xi _{k}\simeq k^{\alpha }  \label{74}
\end{equation}
Then, for small velocities, the populations are proportional to: 
\begin{equation}
X_{k}\simeq k^{-\alpha }  \label{75}
\end{equation}
Now insert these expression into expressions (\ref{72}) and (\ref{73});
since $\delta \xi _{k}$ is given by a 6 dimension integral of three
quantities which vary as $k^{-\alpha }$, we get the relation: 
\begin{equation}
k^{6-3\alpha }=k^{\alpha }  \label{76}
\end{equation}
or: 
\begin{equation}
\alpha =\frac{3}{2}  \label{77}
\end{equation}
In the above reasoning, we have implicitly assumed that the behavior of $%
\delta \xi _{k}$ for small values of $k$ depends only on the domain of
integration close to the origin; our result is then consistent since a
function in $k^{3/2}$ will dominate over a quadratic function as long as $k$
is sufficiently small.

As far as the second approximation is concerned, the simplest possibility to
go beyond it is to include more diagrams and to use (\ref{57}) instead of (%
\ref{56}), which amounts to adding a denominator $1+J(\mathbf{k}+\mathbf{k}%
^{^{\prime }})$ under $J(\mathbf{k}+\mathbf{k}^{^{\prime }})$ inside the
integral (\ref{72}). The analytical algebra then becomes even more
complicated, but we have performed the corresponding numerical
calculations.\ The results are shown in the middle line in figure 14:
indeed, they are in better agreement with the quantum Monte-Carlo results
than the more crude approximation made before. One could therefore hope that
a really good agreement would be attainable by including a larger class of
diagrams in our calculations. It is also interesting to note that our
conclusion concerning the exponent $\alpha =3/2$ remains valid in this case,
as actually also when a broader class of diagrams is added in the integral
equation for $\rho _{1}$; this is because more and more diagrams can be
generated in a process where one $U_{2}$ is added, bringing with it two $X$%
's, but also one momentum integration, a process in which the low $k$
dependence of $\delta \xi _{0}-\delta \xi _{k}$ is multiplied by $%
k^{3-2\alpha }$, a constant if $\alpha =3/2$.

\section{Conclusion}

Our study show that the effect of repulsive interactions is to shift the
degeneracy parameter by an amount: 
\begin{equation}
\frac{\Delta (n\lambda _{T})_{cr.}}{(n\lambda _{T})_{cr.}}\simeq -1.3\,\,%
\frac{a}{\lambda _{T}}\simeq -(na^{3})^{1/3}  \label{concl1}
\end{equation}
or, in terms of a shift of critical temperature of the gas at constant
density: 
\begin{equation}
\frac{\Delta T_{c}}{T_{c}}\simeq \frac{a}{\lambda _{T}}\simeq
0.7\,\,(na^{3})^{1/3}  \label{concl2}
\end{equation}
where $a$ is the scattering length (assumed to be positive) of the
interaction potential and $\lambda _{T}$ the thermal wavelength of the
atoms.\ In the case of hard cores, these results are about 100\% larger than
those resulting from the numerical calculations of \cite{Gruter}, which
provides a reasonable agreement in view of the approximations that we have
made when solving the final non-linear equations; in fact, it is perfectly
possible that more extensive calculations would lead to an even closer
agreement. A striking feature of the results of \cite{Gruter} is the
presence of two regimes, due to the increase of $T_{c}$ at low densities
introduced by the repulsive interactions, followed by a decrease at higher
densities. This is not an obvious phenomenon: usually interactions tend to
mask quantum effects and not to enhance them, so that one could naturally
expect a constant decrease of the transition temperature as a function of
density.\ Our approach explains this feature naturally, in terms of a
microscopic re-arrangement mechanism which, in essence, automatically leads
to an increase of this temperature (for repulsive interactions); the
re-arrangement produces an enhancement of the velocity distribution at its
center, which develops just above the transition temperature as a precursor,
and allow the gas to reach the transition point before its density
(integrated over all velocities) reaches the critical density of the ideal
gas. In this picture, the mechanism for superfluidity is inherently
different from Bose-Einstein condensation in an ideal gas, and in a sense
one could say that superfluidity triggers condensation at a lower density
(or higher temperature) than in a mean field picture. At the transition
point, the distortion of the velocity distribution bears some similarity
with the results of a Bogolubov transformation (where significant changes of
the populations are also introduced for low energy levels), but the exponent
is different, with an intermediate value between the high temperature and
low temperature limit; in a future article, we intend to extend our theory
to lower temperatures and condensed gases in order to make a more precise
contact with this transformation.\ One can also remark that the singular
behavior of the mean-field theory near the transition point introduces
non-analyticities in the results, mixing up together various orders in the
natural expansion parameter $a/\lambda _{T}$; the presence of a square root
in the calculations makes it necessary to push the calculation to higher
orders than one would naively expect.\ This suggest that theories using
pseudopotentials may quickly reach their limit, since they are based on
expressions of the matrix elements which are valid up to the second power of
the potential range; from third order, the behavior of wave functions inside
the interaction potential (particles in the middle of a collision) becomes
relevant, so that an Ursell approach would be safer.

Concerning the high density regime, it remains for the moment beyond the
domain of validity of our calculations: in denser systems, a treatment in
terms of Ursell operators or rank exceeding 2 would become necessary.\ Doing
so, one could then hope to see how a mean attractive field develops in order
to stabilize a liquid, even before Bose-Einstein condensation is reached -
in other words, develop a more complete theory predicting the existence of
two transitions, ordinary liquefaction and superfluid transition (as in
helium four).\ But denser systems do not only differ from gases because of
the presence of an attractive mean field; the atomic motions are also
hindered by the interactions for lack of space, so that the spatial
rearrangement which can easily take place in a gas is no longer possible.\
The physics of the effects of the interactions on exchange and on the
critical temperature of Bose-Einstein transition is therefore significantly
different. In a similar perspective, we note that several of the integrals
that have been introduced in the treatment beyond the Ursell-Dyson
approximation are mathematically different from those of a mean-field
treatment: higher dimensions integrals, etc..\ In other words, the
dimensional properties of the transition may be significantly altered by the
interactions; this is another possibility that should be explored.

Another approximation which could be released is ignoring the changes as a
function of the relative momentum of the Ursell length (or the scattering
length), which we have treated as exactly constant; moreover, we have not
included the energy dependence of the off-diagonal matrix elements of $U_{2}$%
.\ A better treatment could for instance include a quadratic dependence on
the relative momentum; one could then reasonably expect to recover the
effective mass effect predicted in \cite{Fetter-Walecka}, since it is
precisely due to a $k$ dependence of the matrix elements, and probably a
better approximation to the full density variations predicted in \cite
{Gruter}. Finally, it could also be interesting to explore the effects of
resonant changes of the cross sections at very low energies on the
transition temperature, since they seem to be now accessible \cite
{aresonant1}\cite{aresonant2}\cite{Ketterl}. One could imagine situations
where sharp variations of the cross section could completely change the
character of the Bose-Einstein transition, maybe even to a point where
condensation would not involve a single level but, for instance, all levels
with a given length of $k$; this would be an even more unconventional
situation than the smeared condensation mentioned in the introduction.

\begin{center}
ACKNOWLEDGMENTS
\end{center}

The preliminary stages of this work were initiated during visits to the
Kamerlingh Onnes Laboratorium in Leiden and the Blackett Laboratory at
Imperial College (London).\ The hospitality of Profs.\ G.\ Frossati, J.P.\
Connerade, H.\ Hutchinson and N.\ Rivier played an essential role in making
these visits fruitful. Discussions with Philippe Nozi\`{e}res, Sandro
Stringari, Alex Meyerovich, Antony Leggett and Edouard Br\'{e}zin were
especially useful. The final part of this work was made during a very
stimulating visit at the Institute of Theoretical Physics at the University
of California at Santa Barbara.

This research was supported in part by the National Foundation under Grant
No.\ PHY94-07194.

\begin{center}
APPENDIX A: expansion of the solution of an integral equation
\end{center}

In this appendix, we study the solution of equation obtained by adding (\ref
{22}) to the right hand side of (\ref{19}): 
\begin{equation}
\rho _{1}(1)=f_{1}(1)+2xz^{2}\,\left[ 1+f_{1}(1)\right] Tr_{2}\left\{ W(1,2)+%
\left[ W(1,2)\right] ^{2}\right\}  \label{a1}
\end{equation}
(for convenience, a variable $x$ has been introduced in front of the trace
but, at the end of the calculation, we can set $x=1$) where: 
\begin{equation}
W(1,2)=U_{2}^{S}(1,2)\left[ 1+\rho _{1}(1)\right] \left[ 1+\rho _{1}(2)%
\right]  \label{a2}
\end{equation}
We show, by a series expansion in powers of $x$, that this integral equation
gives correct weights (in fact, weights equal to one) to all diagrams that
it introduces; a similar proof could be made for the more general equation (%
\ref{23}). If we expand the solution of (\ref{a1}) in powers of $x$: 
\begin{equation}
\rho _{1}(1)=\sum_{n=0}^{\infty }x^{n}\rho _{1}^{(n)}(1)  \label{a3}
\end{equation}
we get from (\ref{a1}) the following recurrence relation: 
\begin{equation}
\begin{array}{lr}
\rho _{1}^{(n)}(1) & \displaystyle=2z^{2}\left[ 1+f_{1}(1)\right]
Tr_{2}\left\{ U_{2}^{S}(1,2)\sum_{q=0}\left[ \delta _{q,0}+\rho _{1}^{(q)}(1)%
\right] \left[ \delta _{n-q,0}+\rho _{1}^{(n-q-1)}(2)\right] \right. \\ 
& \displaystyle+U_{2}^{S}(1,2)\sum_{q+q^{^{\prime }}+q^{^{\prime \prime
}}+q^{^{\prime \prime \prime }}=n-1}\left[ \delta _{q,0}+\rho _{1}^{(q)}(1)%
\right] \left[ \delta _{q^{^{\prime }},0}+\rho _{1}^{(q^{^{\prime }})}(2)%
\right] \\ 
& \displaystyle\left. U_{2}^{S}(1,2)\left[ \delta _{q^{^{^{\prime \prime
}}},0}+\rho _{1}^{(q^{^{^{\prime \prime }}})}(1)\right] \left[ \delta
_{q^{^{\prime \prime \prime }},0}+\rho _{1}^{(q^{^{\prime \prime \prime
}})}(2)\right] ^{\stackrel{}{}}\right\}
\end{array}
\label{a4}
\end{equation}
for $n\neq 0$; for $n=0$ we get: 
\begin{equation}
\rho _{1}^{(0)}(1)=f_{1}(1)  \label{a5}
\end{equation}
Although equation (\ref{a4}) looks complicated, its content is actually
simple; it describes how the structure of the diagrams contained in $\rho
_{1}^{(n)}(1)$ is built from the ``root'' of the diagram, progressing along
the lowest horizontal cycle/line, and adding branches made of lower orders.\
For example, let us first discuss direct terms, those for which $U_{2}^{S}$
can be merely replaced by $U_{2}$. When the first $U_{2}$ operator occurs
along the lowest horizontal line, two cases are possible: either this
operator introduces a new horizontal line (cycle) which is not connected
anymore to the root line by any other $U_{2}$, and this case corresponds to
the first line in the right hand side of (\ref{a4})); or it introduces a new
horizontal line which is connected again by another $U_{2}$ to the root
line, corresponding to the second and third line in (\ref{a4}).\ In the
first case, two sub-diagrams are branched after this first $U_{2}$, as shown
by the arrows in the fist diagram of fig.16; they can occur in $n$ possible
ways (this corresponds to the summation over $q$), combining all possible
lower orders to get a sum of orders equal to $n-1$; first, the lower
connection is replaced by a dotted line symbolizing $1+f_{1}(1)$ (the sum of 
$U_{1}(1)$ to all powers ranging from $1$ to infinity) while the upper line
contains the diagrams contained in $\rho _{1}^{(n-1)}$; second, the lower
connection gets $\rho _{1}^{(1)}$ while the upper connection gets $\rho
_{1}^{(n-2)}$, etc.. In the second case, the basic idea is the same, except
that now the lower orders $\rho _{1}^{(q)}$ are plugged in at four different
places, as symbolized in the second diagram of figure 16.\FRAME{dtbpFU}{%
2.1672in}{2.2952in}{0pt}{\Qcb{Fig.\ 16: Diagrams summarizing the
construction of $\protect\rho _{1}^{(n)}$ as a function of $\protect\rho 
_{1}^{(n-1)}$.}}{}{fg16.eps}{\special{language "Scientific Word";type
"GRAPHIC";maintain-aspect-ratio TRUE;display "USEDEF";valid_file "F";width
2.1672in;height 2.2952in;depth 0pt;original-width 4.7496in;original-height
5.0306in;cropleft "0";croptop "1";cropright "1";cropbottom "0";filename
'fg16.eps';file-properties "XNPEU";}}

As for the exchange terms, where $U_{2}^{S}$ is replaced by $%
U_{2}^{S}P_{ex.} $, they are very similar, except that the first $U_{2}$
reconnects to the root cycle, as shown in the third and fourth diagrams in
figure 16.\ Finally, direct and exchange connections can be combined in any
way so that, eventually, all diagrams corresponding to the class considered
are generated.\ By recurrence we see that, if any direct Ursell diagram was
contained once and only once in $\rho _{1}^{(n-1)}$, it will also be
contained once and once only at order $n$ in $\rho _{1}^{(n)}$. The integral
equation therefore ascribes weights to all diagrams which are 1, as
expected. It therefore gives an appropriate description of the approximation
considered, where two cycles are never connected more than twice.

The generalization to any number of connections is the purpose of equation (%
\ref{23}); it can bee checked in the same way that it corresponds to
appropriate weights for all the diagrams.

\begin{center}
APPENDIX B: a potential minimization solution of the non linear equations
\end{center}

Define the potential: 
\begin{equation}
\Phi =\frac{1}{2}\sum_{\mathbf{k},\mathbf{k}^{^{\prime }}}\alpha _{\mathbf{k}%
,\mathbf{k}^{^{\prime }}}X_{\mathbf{k}}X_{\mathbf{k}^{^{\prime }}}-\sum_{%
\mathbf{k}}\left( \xi _{\mathbf{k}}X_{\mathbf{k}}+\log X_{\mathbf{k}}\right)
\label{c1}
\end{equation}
It is stationary under the conditions: 
\begin{equation}
\frac{\partial \Phi }{\partial X_{\mathbf{k}}}=\sum_{\mathbf{k}^{^{\prime
}}}\alpha _{\mathbf{k},\mathbf{k}^{^{\prime }}}X_{\mathbf{k}^{^{\prime
}}}-\xi _{\mathbf{k}}-\frac{1}{X_{\mathbf{k}}}=0  \label{c2}
\end{equation}
which, assuming $X_{\mathbf{k}}\neq 0$, is equivalent to (\ref{34})
(actually to a more general version of these equations where all $\alpha _{%
\mathbf{k},\mathbf{k}^{^{\prime }}}$ are not necessarily equal). The
solution of the non-linear equations of the Ursell-Dyson approximation can
therefore be obtained by a potential minimization of the function $\Phi $.

Assume for instance that all $X_{\mathbf{k}}$'s except two, $X_{0}$ and $%
X_{1}$, are kept constant.\ We set: 
\begin{equation}
\begin{array}{l}
S=X_{0}+X_{1} \\ 
D=X_{0}-X_{1}
\end{array}
\label{c2bis}
\end{equation}
and:

\begin{equation}
\begin{array}{l}
\xi _{0}^{^{\prime }}=\xi +\lambda \\ 
\xi _{1}^{^{\prime }}=\xi +\lambda
\end{array}
\label{c3}
\end{equation}
so that the potential function becomes, within terms which do not depend on $%
X_{0}$ and $X_{1}$(from now on we assume that all $\alpha $'s are equal): 
\begin{equation}
\Phi =\frac{\alpha }{2}S^{2}-\xi S-\lambda D-\log \left[ S^{2}-D^{2}\right]
+...  \label{c4}
\end{equation}

We now write: 
\begin{equation}
\begin{array}{l}
\displaystyle\frac{\partial \Phi }{\partial S}=\alpha S-\xi -\frac{2S}{%
S^{2}-D^{2}} \\ 
\displaystyle\frac{\partial \Phi }{\partial D}=-\lambda +\frac{2S}{%
S^{2}-D^{2}}
\end{array}
\label{c5}
\end{equation}
The second equation provides a quadratic equation in $D$ with the solution: 
\begin{equation}
D=\frac{1}{\lambda }\left[ -1\pm \sqrt{1+\lambda ^{2}S^{2}}\right]
\label{c6}
\end{equation}
but it is easy to see that only the root with the $+$ sign in front of the
radical is acceptable: in (\ref{30}), the $X_{\mathbf{k}}$'s are defined as
positive quantities, and the other solution would lead to $D<S$. For a
system contained in a box of size $L$, the energy difference between the
ground state and the first excited level is proportional to $L^{-2}$, so
that the same is true for $\lambda $; now, if we assume that the sum of the
populations $S$ is macroscopic, we have $S\sim L^{3}$ and $\lambda S\gg 1$.\
Then: 
\begin{equation}
D\simeq \frac{1}{\lambda }\left[ \lambda S\left( 1+\frac{1}{2(\lambda S)^{2}}%
\right) -1\right] =S-\frac{1}{\lambda }  \label{c7}
\end{equation}
which is, as $S$, proportional to $L^{3}$.\ We see that: 
\begin{equation}
\begin{array}{l}
X_{1}=(S-D)/2\sim L^{2} \\ 
X_{0}=(S+D)/2\sim L^{3}
\end{array}
\label{c7bis}
\end{equation}
which is similar to the situation for the ideal gas. Finally, $S$ is given
by the first equation (\ref{c5}): 
\begin{equation}
\frac{\partial \Phi }{\partial S}=0=\alpha S-\xi -\frac{2S}{2D/\lambda }%
=\alpha S-\xi -\lambda  \label{c8}
\end{equation}
which provides: 
\begin{equation}
S\simeq \frac{\xi _{1}^{^{\prime }}}{\alpha }\sim \mathcal{V}  \label{c9}
\end{equation}
The advantage of the potential method is that it allows one to go beyond the
continuous (integral) approximation of $\Delta \overline{\xi }$ and to study
what happens to each discrete level; it would be interesting to generalize
the method to equations (\ref{56}) and (\ref{57}) in order to check the
quality of the conjectures made in \S\ \ref{discussion} concerning the
linear character of the velocity profile near the center.

\begin{center}
APPENDIX C: matrix elements of $U_{2}$.
\end{center}

In this appendix we give a perturbative calculation of the matrix elements
of $U_{2}$; for a strong potential of range $b$, replacing the real
potential by a pseudopotential \cite{Huang-1} \cite{Huang-et-coll.}, the
calculation remains valid up to second order in $b$. The first order value
of the part of $U_{2}$ which acts in the space of relative motion of the two
particles is: 
\begin{equation}
U_{2}^{rel.}=-\int_{0}^{\beta }d\beta ^{^{\prime }}\text{e}^{-(\beta -\beta
^{^{\prime }})H_{0}}\,\,V_{2}\,\,\text{e}^{-\beta ^{^{\prime }}H_{0}}+...
\label{b1}
\end{equation}
where $H_{0}$ is the kinetic energy hamiltonian. From this we get, still in
the space or the relative motion: 
\begin{equation}
<\mathbf{\varkappa }\mid U_{2}^{rel.}\mid \mathbf{\varkappa }^{^{\prime }}>=%
\overline{V}_{2}(\mathbf{\varkappa }^{^{\prime }}-\mathbf{\varkappa })\,\,\,%
\frac{\text{e}^{-\beta \epsilon (\varkappa ^{^{\prime }})}-\text{e}^{-\beta
\epsilon (\varkappa )}}{\epsilon (\varkappa ^{^{\prime }})-\epsilon
(\varkappa )}+...  \label{b2}
\end{equation}
where: 
\begin{equation}
\epsilon (\varkappa )=\frac{\hbar ^{2}\varkappa ^{2}}{m}  \label{b2bis}
\end{equation}
is the kinetic energy and $\overline{V}_{2}(\mathbf{k})$ the Fourier
transform of the interaction potential $V_{2}(\mathbf{r})$. Coming back to
the full space of two interacting particles, we set: 
\begin{equation}
\begin{array}{ll}
\displaystyle\mathbf{k}_{1}=\frac{\mathbf{K}}{2}+\mathbf{\varkappa } & %
\displaystyle\mathbf{k}_{1}^{^{\prime }}=\frac{\mathbf{K}}{2}+\mathbf{%
\varkappa }^{^{\prime }} \\ 
\displaystyle\mathbf{k}_{2}=\frac{\mathbf{K}}{2}-\mathbf{\varkappa } & %
\displaystyle\mathbf{k}_{2}^{^{\prime }}=\frac{\mathbf{K}}{2}-\mathbf{%
\varkappa }^{^{\prime }}
\end{array}
\label{b3}
\end{equation}
(if $\varkappa \neq \varkappa ^{^{\prime }}$, the kinetic energy is not
conserved for the relative motion, while it is always for the center of mass
motion; in the same way, the total momentum is always conserved). We the
have: 
\begin{equation}
<\mathbf{k}_{1},\mathbf{k}_{2}\mid U_{2}(1,2)\mid \mathbf{k}_{1}^{^{\prime
}},\mathbf{k}_{2}^{^{\prime }}>=\text{e}^{-\beta E_{K}}\overline{V}_{2}(%
\mathbf{q})\frac{\text{e}^{-\beta \epsilon (\varkappa ^{^{\prime }})}-\text{e%
}^{-\beta \epsilon (\varkappa )}}{\epsilon (\varkappa ^{^{\prime
}})-\epsilon (\varkappa )}+...  \label{b4}
\end{equation}
where $E_{K}$ is the kinetic energy of the center of mass: 
\begin{equation}
E_{K}=\frac{\hbar ^{2}K^{2}}{4m}  \label{b4b}
\end{equation}
and $\mathbf{q}$ is the transferred momentum: 
\begin{equation}
\mathbf{q}=\mathbf{\varkappa }-\mathbf{\varkappa }^{^{\prime }}
\label{b4bis}
\end{equation}
\ With this result, if we set: 
\begin{equation}
z^{2}<\mathbf{k},\mathbf{k}^{^{\prime }}\mid U_{2}^{S}\mid \mathbf{k}+%
\mathbf{q},\mathbf{k}^{^{\prime }}-\mathbf{q}>=-4\frac{\lambda _{T}^{2}}{%
\mathcal{V}}\,\,\text{e}^{-\frac{\beta }{2}\left[ \widetilde{e}_{\mathbf{k}}+%
\widetilde{e}_{\mathbf{k}^{^{\prime }}}+\widetilde{e}_{\mathbf{k}+\mathbf{q}%
}+\widetilde{e}_{\mathbf{k}^{^{\prime }}-\mathbf{q}}\right] }\times a(\text{%
non-diagonal})  \label{b5}
\end{equation}
we get: 
\begin{equation}
a(\text{non-diagonal})=\overline{V}_{2}(\mathbf{q})\frac{\sinh \beta \left[ 
\widetilde{e}(\varkappa ^{^{\prime }})-\widetilde{e}(\varkappa )\right] }{%
\beta \left[ \widetilde{e}(\varkappa ^{^{\prime }})-\widetilde{e}(\varkappa )%
\right] }  \label{b6}
\end{equation}
which shows that the off diagonal matrix elements remain comparable to the
diagonal elements within the thermal profile, with a factor of approximately
2.

\begin{center}
APPENDIX\ D: calculating the correction to the chemical potential introduced
by velocity-dependent effects.
\end{center}

Equations (\ref{49bis}) and (\ref{50}), together with the definition (\ref
{46}) of $\Delta \overline{\mu }$, provide the relation: 
\begin{equation}
\Delta \overline{\xi }=\frac{4a}{\lambda _{T}}\,\,\frac{1}{1-\Delta 
\overline{\xi }}\,\,g_{3/2}\left[ z\left( 1-\Delta \overline{\xi }\right) %
\right]  \label{d1}
\end{equation}
which can also be written as: 
\begin{equation}
\text{e}^{-\beta \Delta \overline{\mu }}-\text{e}^{-2\beta \Delta \overline{%
\mu }}=\frac{4a}{\lambda _{T}}\,\,\,\,g_{3/2}\left[ \text{e}^{\beta (\mu
-\Delta \overline{\mu })}\right]  \label{d2}
\end{equation}
Writing the same relation for the particular value of $\mu $: 
\begin{equation}
\mu =\overline{\mu }_{cr.}=\Delta \overline{\mu }_{cr.}  \label{d3}
\end{equation}
provides: 
\begin{equation}
\text{e}^{-\beta \Delta \overline{\mu }_{c}}-\text{e}^{-2\beta \Delta 
\overline{\mu }_{c}}=\frac{4a}{\lambda _{T}}\,\,\,\,g_{\max .}  \label{d4}
\end{equation}
where: 
\begin{equation}
g_{\max .}=g_{3/2}(1)=2.612...  \label{d5}
\end{equation}

Now, by difference, we obtain: 
\begin{equation}
\begin{array}{cc}
\displaystyle\text{e}^{-\beta \Delta \overline{\mu }}-\text{e}^{-\beta
\Delta \overline{\mu }_{c}}-\text{e}^{-2\beta \Delta \overline{\mu }}+\text{e%
}^{-2\beta \Delta \overline{\mu }_{c}} & \displaystyle=\frac{4a}{\lambda _{T}%
}\left[ g_{3/2}\left[ \text{e}^{\beta (\mu -\Delta \overline{\mu })}\right]
-g_{\max .}\right] \\ 
& \displaystyle\simeq -\frac{4a}{\lambda _{T}}c\sqrt{\beta \left( \mu
\right) -\Delta \overline{\mu }}
\end{array}
\label{d6}
\end{equation}
where the numerical coefficient $c$ is given by (see \cite{Robinson} or
exercise 12.3 of \cite{Huang-1}): 
\begin{equation}
c=3.544..  \label{d7}
\end{equation}
From this we obtain: 
\begin{equation}
\beta \left( \mu -\Delta \overline{\mu }\right) =\left( \frac{\lambda _{T}}{%
4a}\right) ^{2}c^{-2}\left[ \text{e}^{-\beta \Delta \overline{\mu }}-\text{e}%
^{-\beta \Delta \overline{\mu }_{c}}-\text{e}^{-2\beta \Delta \overline{\mu }%
}+\text{e}^{-2\beta \Delta \overline{\mu }_{c}}\right] ^{2}  \label{d8}
\end{equation}
or: 
\begin{equation}
\beta \left( \mu -\Delta \overline{\mu }\right) =\left( \frac{\lambda _{T}}{%
4a}\right) ^{2}c^{-2}\left[ \beta \left( \Delta \overline{\mu }-\Delta 
\overline{\mu }_{c}\right) \right] ^{2}+...  \label{d9}
\end{equation}
But, to this order, we can replace $\Delta \overline{\mu }$ by $\mu $ in the
right hand side: 
\begin{equation}
\beta \left( \mu -\Delta \overline{\mu }\right) =\left( \frac{\lambda _{T}}{%
4a}\right) ^{2}c^{-2}\left[ \beta \left( \mu -\Delta \overline{\mu }%
_{c}\right) \right] ^{2}+...  \label{d10}
\end{equation}

At the critical point, $\mu +\delta \mu _{0}-\Delta \overline{\mu }=0$ so
that this relation becomes: 
\begin{equation}
\beta \delta \mu _{0}\simeq \left( \frac{\lambda _{T}}{4a}\right) ^{2}c^{-2}%
\left[ \beta \left( \mu -\Delta \overline{\mu }_{c}\right) \right] ^{2}+...
\label{d11}
\end{equation}
or: 
\begin{equation}
\beta \left( \overline{\mu }_{c}-\mu _{c}\right) \simeq \frac{4a}{\lambda
_{T}}c\sqrt{\beta \delta \mu _{0}}  \label{d12}
\end{equation}
which is second order in $a/\lambda _{T}$\medskip\ (since $\delta \mu _{0}$
is second order in this quantity).

\end{document}